\documentclass{elsarticle}

\usepackage{lineno,hyperref}
\modulolinenumbers[5]

\journal{Astronomy and Computing}

\usepackage{graphicx}
\usepackage{amsmath}
\usepackage{amssymb}
\usepackage{aas_macros}

\begin{document}

\begin{frontmatter}


\title{On the effect of projections on convergence peak counts and Minkowski functionals}
\author{Z. M. Vallis$^a$\corref{author1}}
\ead{zoe.vallis.14@ucl.ac.uk}
\author{C. G. R. Wallis$^a$}
\author{T. D. Kitching$^a$}

\address{$^a$Mullard Space Science Laboratory, University College London, Holmbury St Mary, Dorking RH5 6NT, UK}
\cortext[author1]{Corresponding author}


\begin{abstract}
The act of projecting data sampled on the surface of the celestial sphere onto a regular grid on the plane can introduce error and a loss of information. This paper  evaluates the effects of different planar projections on non-Gaussian statistics of weak lensing convergence maps. In particular we investigate the effect of projection on peak counts and Minkowski Functionals (MFs) derived from convergence maps and the suitability of a number of projections at matching the peak counts and MFs obtained from a sphere. We find that the peak counts derived from planar projections consistently overestimate the counts at low SNR thresholds and underestimate at high SNR thresholds across the projections evaluated, although the difference is reduced when smoothing of the maps is increased. In the case of the Minkowski Functionals, $V_0$ is minimally affected by projection used, while projected $V_1$ and $V_2$ are consistently overestimated with respect to the spherical case. 
\end{abstract}



\end{frontmatter}



\section{Introduction}

Gravitational lensing has proven to be a useful tool for probing the cosmology of the Universe and the field is a rapidly growing one, with current surveys providing a wealth of weak lensing data (e.g. the Kilo Degree Survey \citep{DeJong2013}, the Dark Energy Survey \citep{DES2005}) and future surveys (e.g. \emph{Euclid} \citep{Euclid2011}, the Large Synoptic Survey Telescope \citep{LSSTScienceCollaboration2009}, the Wide Field Infrared Survey Telescope \citep{Spergel2013}) promising to further expand the potential of the field. Gravitational lensing occurs when light from distant galaxies passes through mass overdensities, causing a distortion in the galaxy ellipticities detectable by observers. Cosmic shear refers to the weak lensing distortion caused by the the Large Scale Structure (LSS) -- intermediate matter between the galaxies and the observer. The distribution and evolution of the LSS is governed by the cosmological model of the Universe, hence the suitability of weak lensing studies of the LSS for probing cosmology. We can probe the structure of the LSS through cosmic shear to analyse dark energy, evaluate different cosmological models and constrain cosmological parameters; for a recent review see \cite{Kilbinger2015a}.

Among the most commonly used cosmic shear analysis methods are two-point statistics -- the power spectrum or the correlation function -- which probe the second-order properties of the cosmic shear. However the power spectrum and correlation function average over phase information on the sky, and thereby reduce spatial information to a single angular or scale dependence, and cause mixing of angular modes. These statistics do not fully capture the non-Gaussian properties of the lensing matter and perform best on linear scales, where the modes are uncorrelated and grow independently, but loose effectiveness on smaller non-linear scales, though studies have attempted to model the resulting two-point statistics in the non-linear regime and model the non-Gaussian properties with the power spectrum \citep{Takada2009, Pielorz2014}. Furthermore, partial sky coverage and masking of the data cause mixing of the E-modes and B-modes of the cosmic shear \citep{Kilbinger2006}.

The increased precision and sky coverage of upcoming surveys can help to mitigate the drawbacks of these two-point statistics, but will not help to capture the non-Gaussian properties of the LSS. The non-linear and non-Gaussian properties of the LSS can potentially be better analysed by alternative statistics such as the peak counts of the convergence map \citep{Lin2015a} and Minkowski functionals (MFs) \citep{Mecke1994} such as the genus statistics \citep{Matsubara2000}, as well as higher-order spectra \citep{Munshi2012}. 

One useful tool in studying the non-Gaussian statistics is to use the reconstructed convergence map, as it captures the non-Gaussian properties of the underlying matter density distribution \citep{Pires2009a, Berge2010, Vikram2015}. 
The peak count statistic is a straightforward evaluation of the number of detected peaks of the reconstructed convergence map as a function of their SNR. The convergence maps represent the line-of-sight integration of the gravitational lensing effect caused by the underlying matter field, and so their peaks trace the locations of maximum projected overdensities in the lensing structure and are associated with high-density galaxy clusters and halos, and hence are sensitive to the halo mass function and cosmological parameters \citep{Marian2009}. The low SNR peaks are frequently contaminated by noise, so studies seeking solely to identify the locations of true peaks focus on high SNR peaks with SNR $\geq 3$. However, peaks at SNR$~1-2$ contain valuable cosmological information, provided the noise is accounted for \citep{Kratochvil2009}. By binning the observed galaxies by redshift, it is possible to trace the evolution of the LSS over time. Previous studies have measured the effectiveness of peak count statistics at constraining cosmological parameters \citep{Lin2015a,Lin2015b,Peel2016,Kacprzak2016,Lin2017}. For a range of SNR values the peaks can be used as tracers of dark matter halos and structure, and so therefore the non-Gaussianities in the matter distribution \citep{Shirasaki2016}.  

There are a number of approaches to recovering mass maps from a catalogue of measured galaxy shapes. The most commonly used method for analysis of large weak lensing data sets is the Kaiser-Squires method \citep{Kaiser1993}, which is based on the direct inversion of observed shear field to the convergence field. The are a number of well known draw backs to this method including noise growth on small scales, and errors due to survey boundaries. Due to its simplicity the Kaiser-Squires
method is the standard method to recover mass-maps from data on large scales. For example,
 the Cosmic Evolution Survey (COSMOS;
\citep{scoville:2007}), the Canada-France-Hawaii Telescope
Lensing Survey (CFHTLenS; \citep{heymans:2012})
and the Dark Energy Survey (DES; \citep{flaugher:2015}) Science Verification (SV) data \citep[respectively,][]{massey:2007,van_waerbeke:2013,chang:2015} all use this method.
Other mass mapping techniques have also
been developed. On the galaxy cluster scale
parametric models \citep[e.g.][]{jullo:2007}) and non-parametric methods \citep[e.g.][]{massey:2015,lanusse:2016} have been presented. Other methods have been developed for 3D mass map reconstruction, to deal with masking and heavy noise-domination of the shear signal \citep{2003MNRAS.344.1307B, Taylor:2004ca, Massey:2003xd, 2009MNRAS.399...48S, 2011ApJ...727..118V, Leonard2014, 2012A&A...539A..85L, 2013A&A...560A..33S}. Recently peak counts have been used to measure cosmological parameters from the DES \cite{2016MNRAS.463.3653K}, where a planar projected mass map was used.

The methods discussed above performing reconstruction on the 2D plane all requiring a projection of the shear. \cite{Wallis2017} demonstrated the extension of the commonly used ``Kaiser-Squires'' method to the sphere and performed a comparison with projections, and this method was subsequently applied to DES data by \cite{DES_massmaps:2017}. \cite{Wallis2017} have made their python code \texttt{massmappy}\footnote{\url{http://www.massmappy.org}} publicly available. Along with the publicly availible \texttt{SSHT}\footnote{\url{http://www.spinsht.org}} \citep{McEwen2011}, they have the ability to evaluate the comparative performance of different projections for the purpose of convergence map reconstruction on the 2D surface of the celestial sphere and we build upon their study for the analysis of projection effects on peak count statistics.

The Kaiser-Squires inversion applies to non-locally to transform the dataset as a whole and does not take into account the mask. The common way of dealing with masked data when using the Kaiser-Squires method is to exclude data falling within the masked regions, but this breaks down when the masking covers large areas of the sky, as investigated by \citet{Wallis2017}. We choose to use a simple mask that removes a solid area of the sky, and seek to investigate how it impacts the projected cases compared to the spherical case.

Future surveys will provide greater sky-coverage than before, with the upcoming \emph{Euclid} survey covering 15,000 deg$^2$ \citep{Euclid2011,Amendola:2016saw} and LSST \citep{LSSTScienceCollaboration2009} covering 20,000 deg$^2$, compared to current surveys DES \citep{DES2005} at 5000 deg$^2$ CFHTLenS \citep{Erben2013a} at 154 deg$^2$. As we move towards greater sky coverage in these upcoming surveys, it becomes necessary to fully understand the effects of planar projection and to seek possible alternatives to reconstruction on the plane. These surveys will cover larger areas of the sky, necessitating that we take the sky geometry into account and providing an incentive to move towards analysis directly on the sphere over planar approximations.

Previous analysis of peak counts and MFs have been performed on the projected plane because current surveys cover small areas of sky where a planar approximation holds. However with future surveys covering significantly larger areas of the sky, the geometry of the sky must be accounted for and it is expected that the projections will no longer accurately capture the full information of weak lensing data on the sky. While planar projection analysis has the advantage of being less computationally demanding than performing analysis on the sphere, the movement of research towards analyzing full-sky data necessitates examination of the performance of planar projections used for deriving statistics containing cosmological information compared to the spherical case.

In this paper we investigate the effect of projections on peak count statistics and Minkowski functionals. Peak count statistics rely on accurate mass map reconstructions and minimisation of noise, which is a significant problem at low peak thresholds. Projections to the plane cannot preserve all the features of the spherical map and prioritise accuracy to different map properties, such as preserving angles or relative area. High resolution is required to detect the fine structure of the convergence map and to minimise the merging of closely positioned peaks during projection. It is expected that these factors will influence the peak counts, resulting in distinct differences between projection methods.
The Minkowski functionals \citep{Mecke1994, Schmalzing1997, Munshi2012, Kratochvil2012, Petri2014} and genus statistics \citep{Matsubara2000, Sato2003}, may also be affected by how projections affect the geometry of the shear data. Alternative methods of deriving statistics from the convergence map involve additional data sets such as the CMB-LSS cross-correlation \citep{Pearson2014,Liu2015}.

The focus of this paper is to evaluate how the peak count statistic and Minkowski Functionals are affected for different projections to the plane. This is done by making convergence map reconstructions on the sphere from simulated data based on a CosmoSIS-generated power spectrum \citep{Zuntz2015}.  The peak counts are calculated from the SNR map instead of the convergence map itself, to allow for more natural thresholding. Section \ref{sec:bg} will present a summary the relevant weak lensing background and summarise the projections used. Section \ref{sec:method} will outline the methods used, including the reconstruction of the convergence map and peak detection. We present our findings in Section \ref{sec:results} and draw our conclusions in Section \ref{sec:conclusion}. Further details on implementation and in particular how the convergence maps are smoothed are contained in \ref{appendix:smoothing}.

\section{Background}
\label{sec:bg}
In this paper we are concerned with constructing the spin-0 convergence map $\kappa$ from spin-2 weak lensing shear data $\gamma$, which are measures of the magnification and the shape distortion of the  source light respectively. The convergence and the shear are related to the underlying lensing potential $\Phi$ by 
\begin{eqnarray}
\label{eq:kappa_del}
\kappa &=&\frac{1}{2}[\eth\overline{\eth}+\overline{\eth}\eth]\Phi\nonumber\\
\gamma &=&\frac{1}{2}[\eth\eth]\Phi,\nonumber\\
\end{eqnarray}
where the operators $\eth$ and $\overline{\eth}$ are the spin raising and lowering operators defined as

\begin{eqnarray}
\eth &=& -\sin^s\theta\bigg(\frac{\partial}{\partial\theta}+\frac{i}{sin\theta}\frac{\partial}{\partial\phi}\bigg)\sin^{-s}\theta\nonumber\\
\overline{\eth} &=& -\sin^{-s}\theta\bigg(\frac{\partial}{\partial\theta}-\frac{i}{sin\theta}\frac{\partial}{\partial\phi}\bigg)\sin^s\theta.\nonumber\\
\end{eqnarray}

We make use of the reduced shear, defined as
\begin{equation} 
\label{eq:reducedshear}
g = \frac{\gamma}{1-\kappa}. 
\end{equation}
In generating a mass map one seeks to construct the spatial distribution of $\kappa$ from measured estimates of $g$.

\subsection{Reconstruction on the Plane}
\label{Reconstruction on the Plane}
The method for reconstruction is described in detail in \cite{Wallis2017}, and we provide a brief explanation here. The method used to reconstruct the convergence map is known as Kaiser-Squires 
reconstruction/inversion \citep{Kaiser1993}. Transforming the convergence and shear, as defined in equations 
(\ref{eq:kappa_del}), into Fourier space allows the relation of the convergence and shear. On the plane, provided that the planar approximation holds, this is done by solving the inverse equation
\begin{equation}
\hat{\gamma}(\ell_x,\ell_y) = \mathcal{E}\hat{\kappa}(\ell_x,\ell_y),
\end{equation}
where $\hat{f}$ denotes the Fourier transform of $f$, $\ell_x$,$\ell_y$ are the Fourier coefficients of the spatial coordinates on the plane and
\begin{equation}
\mathcal{E} = \frac{\ell_x^2-\ell_y^2+i2\ell_x\ell_y}{\ell_x^2+\ell_y^2}.
\end{equation}
This is inverted and uses the property $\mathcal{E}^{-1}=\mathcal{E}^*$ to obtain a the Kaiser-Squires estimator for the convergence in Fourier space,
\begin{equation}
\hat{\kappa}(\ell_x,\ell_y) = \mathcal{E}^*\hat{\gamma}(\ell_x,\ell_y).
\label{eq:KSplane}
\end{equation}
Applying the inverse Fourier transform to this $\hat{\kappa}(\ell_x,\ell_y)$ estimator will give the convergence in real space, which can be used to construct a map of convergence from shear observations.

\subsection{Reconstruction on the Sphere}
An analogous equation can be obtained in the spherical case \citep{Wallis2017} by using the spherical harmonic transform to give
\begin{equation}
\hat{\gamma}_{\ell m} = \mathcal{D}_{\ell}\hat{\kappa}_{\ell m},
\end{equation}
where $\ell$ and $m$ are the harmonic coefficients of the spatial coordinates and
\begin{equation}
\mathcal{D}_{\ell} = \frac{-1}{\ell(\ell+1)}\sqrt{\frac{(\ell+2)!}{(\ell-2)!}}.
\end{equation}
The process of reconstructing the convergence map involves inverting this relation 
in spherical harmonic space in a similar method to the Kaiser-Squires estimator on the plane
\begin{equation}
\hat{\kappa}_{\ell m} = \mathcal{D}_{\ell}^{-1}\hat{\gamma}_{\ell m}. 
\label{eq:KSsphere}
\end{equation}
Finally performing the inverse spherical harmonic transform on $\kappa_{\ell m}$ allows the convergence map $\kappa$ to be recovered. 

The reconstruction of the convergence map from simulated shear maps will be performed with the {\tt massmappy} code \citep{Wallis2017} that uses equations (\ref{eq:KSplane}) and (\ref{eq:KSsphere}), as well as for the reconstruction on the plane.

\subsection{Measured Statistics}
The two statistics we investagate are peak statistics and Minkowski functionals. Peak statistics refers to the number of local maxima in the reconstructed maps as a function of signal to noise of the peak; the method for identifying peaks is described in Section \ref{Identifying local peaks}.

\subsubsection{Minkowski functionals}
\label{sec:MF}
In addition to peak count statistics we analyse the effect of projection on the Minkowski functionals. Minkowski functionals provide a statistical measure of the morphological features of random fields. They allow probing of high-order non-Gaussian properties \citep{Schmalzing1997,Munshi2012,Petri2014} arising from the random fluctuations in the shear data. For a 2D random field, we can obtain three Minkowski functionals $V_0$, $V_1$ and $V_2$, which respectively serve as a measure of the area, boundary length and Euler characteristic of the excursion set of the 2D field as a fraction of the total area of the field. We define these Minkowski functionals through the following equations \citep{Petri2014}
\begin{equation}
V_0(\nu_0) = \frac{1}{A}\int\Theta(\nu(\textbf{x})-\nu_0) \textrm{da},
\label{eq:V0}
\end{equation}
\begin{equation}
V_1(\nu_0) = \frac{1}{A}\int\delta(\nu(\textbf{x})-\nu_0)\sqrt{\nu_x^2\nu_y^2} \textrm{da},
\label{eq:V1}
\end{equation}
\begin{equation}
V_2(\nu_0) = \frac{1}{A}\int\delta(\nu(\textbf{x})-\nu_0)\frac{2\nu_x\nu_y\nu_{xy} - \nu_x^2\nu_yy - \nu_y^2\nu_xx}{\nu_x^2\nu_y^2} \textrm{da},
\label{eq:V2}
\end{equation}
where $\Theta$ is the Heaviside step function, $\delta$ is the Dirac delta function, $\nu_x$ and $\nu_y$ denote partial differentiation on the horizontal and vertical (or latitudinal and longitudinal) coordinate directions respectively, $A$ is the total area of the map, $\textrm{da}$ denotes the area element of the map, $\nu$ is the SNR map and $\nu_0$ is the SNR threshold. When integrating over $\textrm{da}$, we integrate along the $x$ and $y$ axes and then divide the result by the total area of the map calculated from the pixel array, as we want the Minkowski functionals expressed as a fraction of the overall area.  We make the Dirac delta functions more appropriate for computation by dividing the range of SNR values into bins $\nu_i$ and using
\begin{equation}
\delta = 
\begin{cases}
1, \quad \text{if } \nu_i-\frac{1}{2}\Delta\nu_i\leq\nu(\textbf{x})\leq\nu_+\frac{1}{2}\Delta\nu_i \\
0, \quad \text{otherwise}
\end{cases}
\end{equation}
where $\nu_i$ is the SNR threshold of bin $i$ and $\Delta\nu_i$ is the width of the SNR threshold bins. For Gaussian random fields, the power spectra can be constructed from the Minkowski functional and vice versa. Non-Gaussian fields require approximations or perturbation methods to relate the power spectra to the Minkowski functionals \citep{Petri2014}.

\subsection{Projections}
\label{sec:proj}
When using standard planar Kaiser-Squires reconstruction (Section \ref{Reconstruction on the Plane}) a projection from the sphere to a plane is required. As discussed in \cite{Wallis2017} such a projection can cause changes in the variance of the reconstructed convergence map, that increase as the angular size of the reconstructed area increases, and is dependent on the projection used. Here we test whether  projections also impact the peak counts and Minkowski funcational measurements obtained from such maps, and to what extent. 

For this study, five projections were used: two equatorial projections (Mercator, sine); and three polar projections (orthographic, stereographic, gnomonic). We describe each projection here. The Mercator projection is given by
\begin{equation}
\begin{split}
& x = \phi - \pi, \\
& y = \text{ln}[\tan(\pi/2-\theta/2)], \\
\label{eq:mercator}
\end{split}
\end{equation}
where $(x$, $y)$ are the planar coordinates, and $(\theta,\phi)$ are the spherical angular coordinates. We define $\theta$ as the polar angle and $\phi$ as the azimuthal angle. The Mercator projection is conformal, so the local angles are preserved in the projection. The Mercator projection does not include the poles. The sinusoidal projection is defined by 
\begin{equation}
\begin{split}
& x = (\phi-\pi)\sin(\theta) \\
& y = \theta.
\label{eq:sine}
\end{split}
\end{equation}
The  sinusoidal projection preserves the relative areas of features on the spherical map, but is not conformal so distorts their shape and affects directionality. 

The polar projections display the map on the plane in polar coordinates, defined on the plane in the form $x=\rho\text{ cos}(\psi)$ and $y=\rho\text{ sin}(\psi)$. The polar projections do not cover the entire sphere in one plane, but instead are split into north and south hemispheres which are centred on the poles. On the southern hemisphere, the parameter $\theta$ for the northern hemisphere is replaced with $\pi-\theta$. The stereographic projection, is a conformal polar projection, and defined by
\begin{equation}
\begin{split}
& \rho = 2\tan\left(\frac{\theta}{2}\right) \\
& \psi = \phi.
\label{eq:SP}
\end{split}
\end{equation}
The orthographic projection projects the sphere as if viewed from an infinite distance and is given by 
\begin{equation}
\begin{split}
& \rho = \sin(\theta) \\
& \psi = \phi.
\label{eq:OP}
\end{split}
\end{equation}
The gnomonic projection is given by 
\begin{equation}
\begin{split}
& \rho = \tan(\theta) \\
& \psi = \phi.
\label{eq:GP}
\end{split}
\end{equation}
The gnomonic projection is a projection that maps points on the surface of the sphere along a line through the centre of the sphere to the tangent plane. As a results maximum viewing angle is limited to be $\theta<\pi/2$, as $\theta=\pi/2$ is at infinity. As the viewing angle approaches $\pi/2$, the shape distortions become significant to the point of obscuring the original map data, hence we selected the viewing angle to be $\theta=\pi/4$. Therefore, the peak counts and calculating the Minkowski Functionals are scaled accordingly to be proportional to the sky coverage of the other projections.

When projecting a spin${-}2$ quantity, for example shear or galaxy ellipticities, one must take account for the local rotations. 
A formalism for calculating these local rotations is outlined in \cite{Wallis2017} for both polar and equatorial projections and is implemented in the projection functions as part of the software package \texttt{SSHT}.

\begin{figure*}
	\centering
	\includegraphics[width=0.75\linewidth]{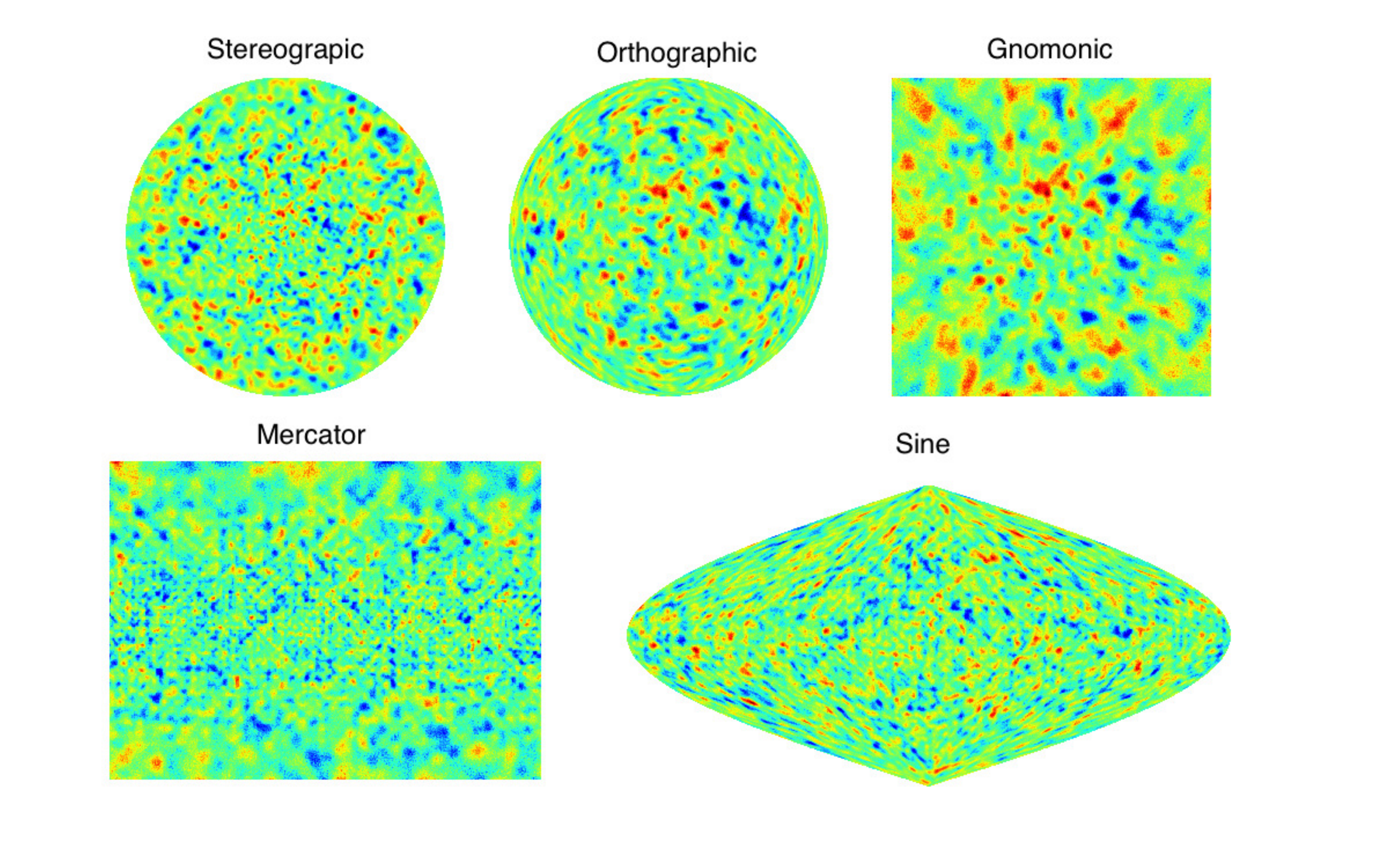}
    \caption{The Signal-to-Noise Ratio map obtained from convergence maps simulated from a template power spectrum, using the method described in Section \ref{sec:convmap} with a standard $\Lambda$CDM cosmology, where the map is defined as $\nu=\frac{\kappa-\overline{\kappa}}{\sigma_{\kappa}}$ for each map. The initial convergence map, which is not displayed, is generated on the sphere. The five displayed convergence maps are the original convergence map projected onto the plane through each of the five projections chosen to be analyzed. This is done by mapping each pixel on the sphere to the corresponding pixel on the planar representation, and taking the mean value when multiple pixels on the sphere are mapped to one pixel on the plane. The center of the projected map is defined as a point on the equator selected to be at the center of the flattened spherical map. From these convergence maps $\kappa$, the SNR maps are defined as $\nu(\bf{\theta})=\frac{\kappa(\bf{\theta})-\overline{\kappa}}{\sigma}$. The maximum bandlimit is selected to be $L=2160$, to match that used in \protect\cite{Vikram2015}, and the resolution for the projections except the Mercator is $2160$, in order to ensure that the projections closely match the spherical map in detail and number of pixels.}
    \label{fig:nu_projections}
\end{figure*}

Each of these projections are displayed in Fig.~\ref{fig:nu_projections}, along with the map on the sphere for the same convergence map. We use the fast and exact sampling theorem of \citep{McEwen2011} to sample the spherical signal, defining upper bandlimit $L$, to define the convergence map on the sphere. The projections are all centered on the same point and the polar projections display only the northern hemisphere. The planar convergence maps were constructed on the planes from the projected shear, allowing one to see the effect of each projection on the reconstruction. We can see that in all cases, the projected image becomes more distorted further from the centre of the projection. Factors that are expected to affect the peak detection include the amount of distortion of the original image and how successfully the noise is smoothed in each projection.

\section{Method}
\label{sec:method}
In this Section we describe the approach taken to assess the impact of any projection on peak count statistics and MFs. The general approach we take is to 1) make simulations of shear fields on the sphere, then 
2) project these using the five selected projections, we then 3) reconstruct the convergence maps either on the sphere, or using planar Kaiser-Squires, and finally 4) measure the peak count statistics and Minkowski functionals in both cases.  

\subsection{Smoothing and Pixelisation}
\label{sec:convmap}
The simulations are produced by generating a simulated power spectrum by adding Gaussian noise to a template cosmic shear power spectrum generated with CosmoSIS \citep{Zuntz2015} using a standard $\Lambda$CDM cosmology $[\Omega_m = 0.3, ~h_0 = 0.72, ~\Omega_b = 0.04, ~\tau = 0.08, ~n_s = 0.96, ~A_s = 2.1e-9, ~\Omega_k = 0.0, ~w = -1.0, ~w_a = 0.0]$. We use \texttt{massmappy} to generate the convergence map in harmonic space from the simulated power spectrum. We then apply smoothing to the convergence signal in harmonic space to mitigate the effect of noise and pixelisation, using Gaussian kernel $\mathcal{G}_l = e^{-l^2\sigma^2}$ where $\sigma= \pi/256$, which provides sufficient but not excessive smoothing. The code \texttt{SSHT} uses the theoretically exact spin spherical harmonic transform with McEwen-Wiaux (MW) sampling \citep{McEwen2011} to transform this harmonic representation of the convergence into the simulated convergence map on the sphere. The simulated reduced shear map is then obtained from this convergence map and random noise, in the form of a Gaussian distribution with $\mu=0$ and $\sigma=1$ multiplied by the standard deviation of the shear map which has average value of $\sim 0.01$, is added to each pixel at this stage. At this point, the convergence map can be recovered and the peak counts calculated on the sphere. To evaluate the projections, the reduced shear data on the sphere is projected into one of the five projections under examination. Following the projection onto the plane, the convergence map reconstruction is performed natively on the plane using standard Kaiser-Squires reconstruction. During the reconstruction step, another Gaussian smoothing is applied to account for noise, with a user-defined $\sigma$. In each case, the smoothing at this step is performed either on the spherical or projected data. We perform the smoothing in the same geometry as that in which the statistical analysis takes place. This reflects what would be done in practice: the projection of the data onto a geometry first in which all subsequent smoothing and data analysis then takes place. Further details on the smoothing during the reconstruction step are discussed in Appendix \ref{appendix:smoothing}.

The maps are processed as pixel arrays, with the dimensions being defined by the maximum bandlimit parameter $L$ for the sphere and the user-defined resolution parameter for the projections. We select the maximum bandlimit to be $L=2160$ to match \cite{Vikram2015}. The number of pixels in the projected maps are defined by the resolution parameter and differ between projections. For each projection, we seek to select resolution parameters that give similar numbers of pixels to the spherical case, to have similar image fidelity and deg$^2$ per pixel. The projection results in a rectangular array of pixels defined by the resolution parameter. However, the sine and orthograpic maps are not rectangular. This results in the number of pixels on these maps being smaller than the total number of pixels produced by the rectangular array that the resolution parameter predicts. We select a resolution parameter equal to $2160$ for the stereographic, orthographic, gnomonic and sine projections.

In order to ensure that the number of pixels for the projections closely matches the number of pixels for the sphere, the resolution was set to be equal to $L$ for all projection methods, except the Mercator projection\footnote{The dimensions of the Mercator projection are defined differently, so we select an equivalent resolution as
$R = \sqrt{N_{sphere}/0.7377}$ where $N_{sphere}$ is the number of pixels in the spherical map, such that and the total number of pixels for the Mercator map is similar to the spherical case. This is because we define the dimensions of the Mercator projection array as $R \times 0.7377 R$, that maps a declination range from $-7\pi/16$ to $7\pi/16$, selected to avoid major projection effects at the poles.}.

The second Gaussian smoothing that occurs when the convergence map $\kappa$ is reconstructed from the shear map $\gamma$ uses a default smoothing of $20$ arcminutes based on the Full-Width Half-Maximum (FWHM), which converts to $\sigma=2\times20.0\times\pi/(60\times180\times2.355) \times \sigma_s$, where $2.355$ is from the FWHM and $\sigma_s$ is a multiplication factor introduced to allow greater ease when discussing adjustments to the smoothing. When the default smoothing based only on the FWHM is used, we set $\sigma_s=1$. We selected $\sigma_s=5$ to allow sufficient smoothing to reduce the effect of the noise, without blurring the convergence map structure. Further details are in \ref{appendix:smoothing}.

The projected maps do not loop around at the boundaries, so therefore the pixels at the edges of the map are not compared against the full 8 pixels they would neighbor on the sphere but instead only the pixels they neighbor in the planar projection. This would result in uncertainty over the validity of such a pixel being a peak, as it is not compared to all of the neighboring pixels as required by the definition of a peak we use. In order to avoid this uncertainty, we do not count any of the boundary pixels as peaks. The boundary pixels are neglected in calculations of the total area of each projection, but used for evaluating their neighboring pixels as peaks. However, the number of peaks on the boundary pixels is negligible compared to the number of non-boundary peaks due to the large overall number of pixels for each map compared to the number of pixels along each boundary. The number of boundary pixels scale $\propto L$, while the total number of pixels in the map scale $\propto L^2$, hence this effect is only prominent at small values of band-limit $L$.

The convergence maps differ in size and number of pixels due to the projection method and resolution selected. The application of masks will also decrease the number of pixels available for peak detection. As a result, the projection and masking will have a significant effect on the peak counts that is not due to the distortion caused by the projection itself. To account for this, we normalise the peak counts by the number of pixels in each map, and to evaluate the relative shapes of the peak counts. The projections are performed to produce maps with similar total numbers of pixels, and therefore similar resolution scales\footnote{Additionally, the gnomonic projection does not cover the full sky and has maximum opening angle of $\theta=\pi/2$, $\pi/4$ on either side of the pole. Therefore it is necessary to scale the gnomonic projection peak counts by an appropriate factor to account for this. The surface area of sky covered by viewing angle from the pole $\frac{\pi}{4}$ is $(\sqrt{2}+2)\pi r^2)$, and the surface area with full sky coverage $\pi$ is $4\pi r^2$. Hence we take the ratio of area with $\frac{\pi}{4}$ opening angle to the full sky area to obtain the factor $(\sqrt{2}+2)/4$ which we divide the raw gnomonic peak counts by to scale to the full sky case. In the masked case, we still apply the projection to the full map and the resolution scale of the gnomonic projection still differs from, so the normalisation is still required.}.

\subsection{Masks}
\label{sec:masks}
The presence of masking must be taken into account when analysing simulated data, in order to more accurately represent real data. In order to evaluate the effect of projection to the 2D plane on masked data, we used a simplified method of applying varying sized rectangular masks. The masks are defined on the sphere, such that the whole sphere is masked except for a circle centered on the defined centerpoint of the map with radius defined by the opening angle. The original mask on the sphere is projected with each of the five projection methods, and then applied to the corresponding projected shear map. The convergence reconstruction is applied after the masking. The projections apply a rotation such that the centre of the unmasked area is at the centre of the projection, and at the centre of the north hermisphere for the polar projections. It can be seen that the degree of distortion is more significant for the polar projections. As the shape distortion increases further from the centre, the unmasked area's shape is more greatly changed and this occurs at a different rate for different projections given the same initial mask on the sphere.

\subsection{Local peak evaluation}
\label{Identifying local peaks}
In order to identify peaks first we need to construct a signal-to-noise (SNR) map. The SNR map is the map containing the ratio of the convergence map at each point to the the standard deviation of the convergence map. We apply peak finding to the SNR map in order to more intuitively define the peak thresholds and to account for different maximum values and variance between maps. This also holds the advantage of being able to distinguish true peaks with higher SNR from the lower SNR peaks that will be heavily noise-dominated. The SNR map is constructed from the convergence map as
\begin{equation}
    \nu(\bf{\theta})=\frac{\kappa(\bf{\theta})-\overline{\kappa}}{\sigma},
	\label{eq:numap}
\end{equation}
where $\overline{\kappa}$ and $\sigma$ denote the mean and standard deviation of the convergence map respectively. In the case of the polar projections, where the sphere is projected onto two separate planar maps for the north and south hemispheres, we calculate the mean and standard deviation over both projected hemispheres.
Peak count statistics typically evaluate the counts above a specific SNR threshold over a range SNR thresholds to extract cosmological information \citep{Shirasaki2016}. For a selected SNR threshold, only peaks with values above this threshold are included in the final count. 

A pixel is defined as being at a peak if the SNR value at this pixel is greater than the SNR values in the eight neighboring pixels. This requires the following conditions to be satisfied
\begin{equation}
	\nu(\theta_{xy}) > \nu(\theta_{ij}) \text{ for } \quad
	\begin{cases}
    	x-1<i<x+1; \\
		y-1<j<y+1; \\
    \end{cases}
    \forall (i,j) \neq (x,y)
	\label{eq:ispeak}
\end{equation}
The advantage to this method is that it is straightforward to use and understand, but can result in false detection if the noise is not fully accounted for. This method is highly local considers only a block of 9 pixels to determine if there is a peak or not, so therefore is expected to be less affected by large-scale shape distortions from projection. More sophisticated methods in the literature have used, such as the aperture mass detection \citep{Schneider1996,Dietrich2010,Marian2012} using tangential alignment of the shear map to identify peaks. We choose to use the 8-neighbour peak definition over these other methods because of its ease of use and discuss given the importance of pixelisation and image resolution for projection.

There are several potential drawbacks relating to this definition of a peak. It is relatively easy for noise to be falsely categorized as a peak, which we attempt to reduce by applying appropriate smoothing. We also expect to find that the noise as the SNR threshold is increased, the proportionate effect of noise is reduced as the SNR values increase. It is possible that there will be cases where a distinct peak is spread out over neighboring pixels that have the same value, leading to no pixel being detected as a peak by this definition. Due to the way we evaluate peaks with our code, the calculations are performed to such a precision that this event is unlikely. Should such failures to detect a peak occur, we expect them to occur in low numbers that compared to the total number of peaks. In cases where there are multiple smaller peaks in an area of high SNR, we treat each as separate peaks by this definition, which is valid as we are measuring distinct peaks regardless of height at which they occur. We also seek to minimise these effects by obtaining our results over a significant number of iterations such that any of these spurious peaks or undetected peaks are small in number compared to the real peaks. It is possible for some of these instances to occur as a result of shape distortion from projection, so if a projection leads to a high number of these instances such that there is a distinct impact, then it must be taken into account when evaluating said projection.

\subsection{Minkowski functional evaluation}
\label{sec:MFcalc}
We use equations \ref{eq:V0}, \ref{eq:V1}, \ref{eq:V2} to construct the Minkowski functionals from the SNR map $\nu$. On the projected SNR maps, we simply take the gradient of the map at each point. The spherical case requires us to account for the geometry of the data on the sphere when calculating the partial derivatives, this is done using the following equations \citep{1998astro.ph..3317W}
\begin{flalign}
&\frac{\partial}{\partial\phi}Y_{lm}(\theta,\phi) = imY_{lm}(\theta,\phi),& \\
&\frac{\partial}{\partial\theta}Y_{lm}(\theta,\phi) = -m\cot{\theta}Y_{lm}(\theta,\phi) - \sqrt{l(l+1)-m(m-1)}Y_{lm-1}(\theta,\phi),& \\
&\frac{\partial^2}{\partial\phi^2}Y_{lm}(\theta,\phi) = -m^2Y_{lm}(\theta,\phi),& \\
&\frac{\partial^2}{\partial\theta^2}Y_{lm}(\theta,\phi) = ((l(l+1)-\frac{m^2}{\sin{\theta}^2}+\cot{\theta}\frac{\partial}{\partial\theta})Y_{lm}(\theta,\phi),& \\
&\frac{\partial^2}{\partial\theta\phi}Y_{lm}(\theta,\phi) = (im^2\cot{\theta} - im\sqrt{l(l+1)-m(m-1)}e^{i\phi}) Y_{lm-1}(\theta,\phi).&
\end{flalign}
We use \texttt{HEALPix} \citep{2005ApJ...622..759G} to calculate the derivatives in spherical harmonic space on the sphere for the spherical case. The map generated is converted into a \texttt{HEALPix} map for analysis, and the calculated gradient fields are converted back into a map with equiangular sampling. The band limits are selected such that minimal information is lost in the conversion. In all cases, the integrations are performed on the the pixelised array, hence the total area used is given as the total number of pixels examined. When the data is masked, we evaluate only the unmasked subsection of the map and treat the number of unmasked pixels as the total area.

\section{Results}
\label{sec:results}

\subsection{Peak counts}

\subsubsection{Full Sky Case}

\begin{figure*}
	\centering
	\includegraphics[width=\linewidth]{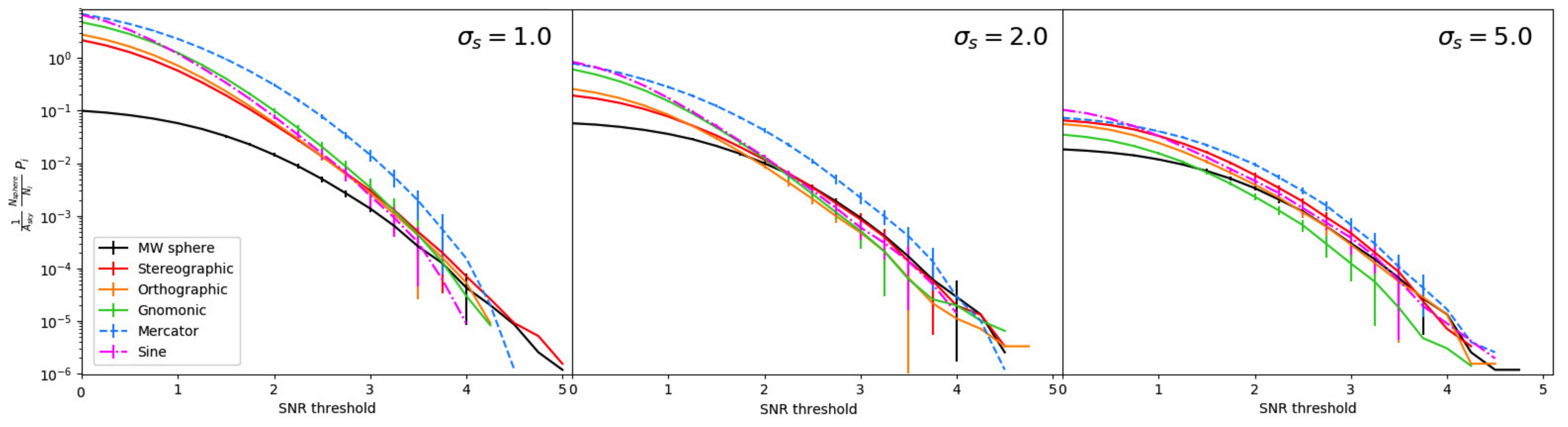}
    \caption{Peak counts divided by covered sky area by SNR threshold for different projections covering the full sky 41253 deg$^2$. The displayed results are the mean of 20 realisations for each projection for the full sky case. We use bandlimit $L=2160$ for the spherical case and corresponding resolutions for each projection. The peak counts are scaled to the area of sky covered and by the ratio of the number of pixels in each projection to the number of pixels on the sphere. The smoothing is $\sigma_s\times 20$ arcminutes on the sky, converted to $\sigma=2\times20.0\sigma_s\times\pi/(60\times180\times2.355)$ in pixel space, where $\sigma_s$ is the Gaussian smoothing scale factor. The three panels show three different Gaussian smoothing scales for the projected maps.}
    \label{fig:peakcount_norm_semilogy}
\end{figure*}

\begin{figure}
	\centering
	\includegraphics[width=0.8\linewidth]{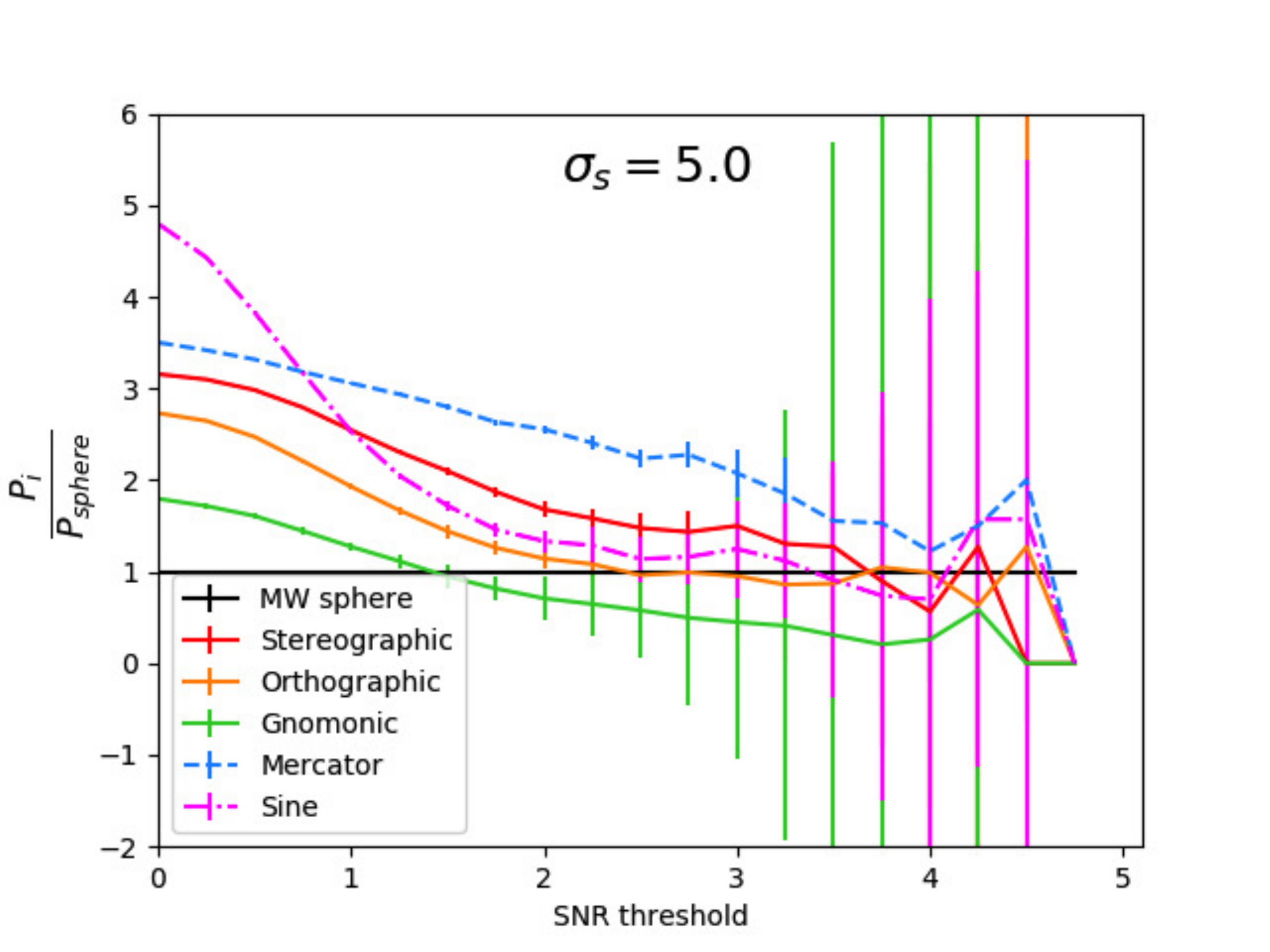}
    \caption{Ratios of the peak counts of SNR thresholds for different projections to the peak counts of the sphere for the full sky case. The peak counts used are the mean of 20 realisations for the unmasked case with smoothing $\sigma_s=5$, corresponding to a Gaussian smoothing factors of $\sigma_s\times 20$ arcminutes on the sky, and have been scaled to the area of sky covered.}
    \label{fig:peakcount_ratios}
\end{figure}

The normalised peak counts for the sphere and the projections are displayed in Fig.~\ref{fig:peakcount_norm_semilogy}. These are the mean values across 10 realisations of randomly simulated shear maps and for maximum bandlimit $L=2160$. The peak counts $P_i$ have been normalised by dividing by the total area of sky covered by each projection, $A_{sky}$, and scaling by ratio of the number of pixels in each projection to the number of pixels in the spherical map, so the peak counts we see are $\frac{N_{\rm sphere}}{N_i}\frac{P_i}{A}$. We can see that the five projection methods produce higher peak counts at low SNR and drop off more rapidly than the spherical case. 

Given that the peak counts are normalised by the area of sky covered for peak identification, the projected peak counts are within one to two order of magnitude of the spherical peak counts. We display the results for three cases of smoothing, with smoothing scales $\sigma_s=1,2,5$ for baseline smoothing of $\sigma_s\times 20$ arcminutes on the sky, and observe that greater smoothing brings the projected peak counts closer to the spherical case peak counts at low SNR thresholds. However, at high SNR thresholds the peak counts are significantly smaller with large errors due to the low number of peaks, so they are often not used for analysis due to unreliability. The SNR thresholds close to SNR$=0$ are also often not used in analysis as they are dominated by noise, so projections that are more accurate to spherical case peak counts on medium SNR thresholds are preferable. In our case, at low SNR thresholds none of the projections closely match the spherical case, although the gnomonic comes the closest. However, since low SNR thresholds are heavily noise-dominated and high SNR thresholds exhibit low peak count numbers, it is more appropriate to compare at moderate SNR thresholds, where we observe that the sine and orthographic projections are reasonably close to the spherical case. It must be noted that the gnomonic projection does not cover the full sky but only the areas up to 45$^{\circ}$ from the poles, and we accommodate this by scaling the peak counts to the reduced area covered.

The ratio of the projected peak counts to the spherical peak counts is displayed in Fig.~\ref{fig:peakcount_ratios} for $\sigma_s=5$. There is a consistent trend across all projections that the number of peaks is overestimated at low SNR threshold. In the threshold range between 1.5 and 3.0, the orthographic and sine projection peak counts are comparable to the spherical peak counts.

The MW-sampled sphere produced by \texttt{ssht} is assumed to be a close approximation of the true underlying data on the celestial sphere. For the case with no masking, the gnomonic projection has the smallest overestimation of the peak count at low SNR thresholds, and is preferred. However the gnomonic projection does not cover the full sky and only the area close to the centre of the projection is undistorted enough to be properly analysed, so it is possible that the extreme distortion far from the centre of projection results in lower peak counts. The peak counts at low SNR thresholds are likely to be dominated by Gaussian noise and cosmological analysis frequently requires detection of higher SNR threshold peaks, so projections that are closer to the spherical case at middle SNR range, such as the orthographic and sine, are useful alternatives for peak count analysis; although we recommend to use the spherical reconstruction.

Examination of the locations of detected peaks on the SNR maps finds that the most common cause for difference in peaks is how projections map features from the sphere to the plane. In cases where more than one pixel on the sphere are mapped to the same projected pixel, the method we used takes the average of these pixels as the value of the projected pixel. Different projection methods may map pixels in a manner such that a pixel that would be identified as a peak in one projection may have a lower value in relation to their neighbours in other projections and so therefore are not defined as a peak. In addition, features on the sphere are scaled differently on the plane for each projection, as can be seen in Fig.~\ref{fig:nu_projections}. This means that for a given area of analysis on the sphere, the corresponding area under analysis on the projected plane will vary in size between projections, leading to the detail being compressed on certain projections compared to others.

\subsubsection{Partial Sky Case}
We examine the partial sky case by finding peak counts for the full sky case and then applying masks of differing sizes, such that we throw away peaks located in the masked area and keep only the peaks in the unmasked area. We assume that boundary effects are negligible. The opening angles are defined as the total angle covered by the unmasked area defined in Section \ref{sec:masks}.

\begin{figure*}
	\centering
	\includegraphics[width=\linewidth]{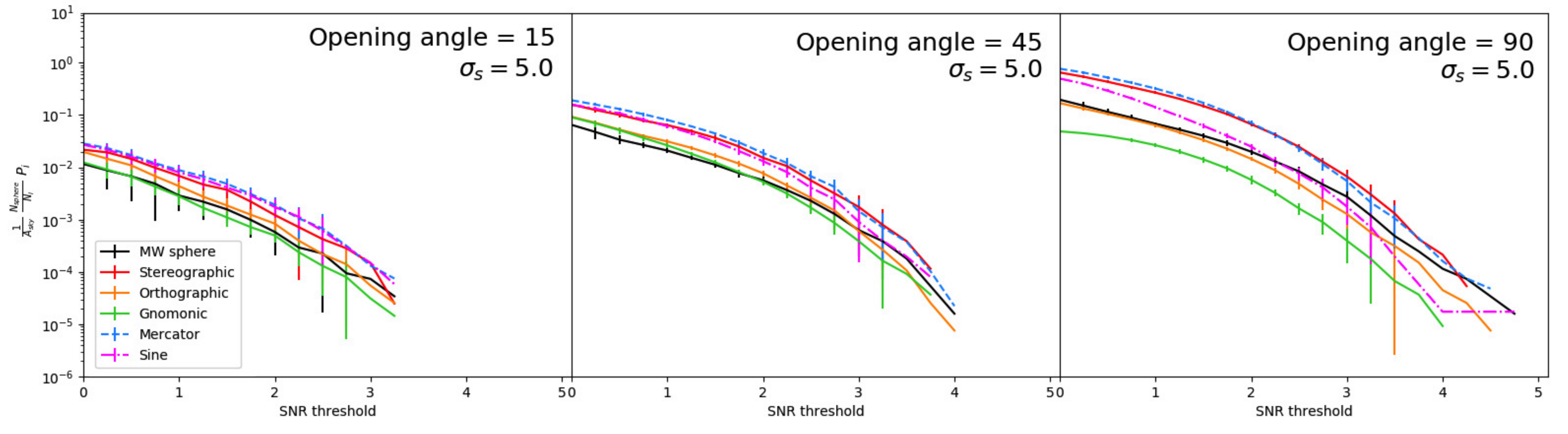}
    \caption{Peak counts divided by covered sky area by SNR threshold for three cases of masked projections covering opening angles 15$^{\circ}$, 45$^{\circ}$ and 90$^{\circ}$. The displayed results are the mean of 20 realisations for each projection and use smoothing scale $\sigma_s=5$ and bandlimit $L=512$ for the spherical case and corresponding resolutions for each projection. The peak counts are scaled to the area of sky covered and by the ratio of the number of pixels in each projection to the number of pixels on the sphere.}
    \label{fig:peakcount_masked}
   \end{figure*}

We can see the overall effect of increasing the unmasked area in Fig.~\ref{fig:peakcount_masked}, where the values are the peak counts normalised in the same way as for the full sky case. When examining the peak count as a function of increasing opening angle for selected SNR thresholds, we find that the peak count grows as a function of sky coverage at approximately the same rate for all projections and the spherical case. The projected peak counts do not converge exactly to the spherical case for any projection, but for certain opening angles and smoothing scales the peak counts of polar projections closely approach the spherical case for SNR thresholds between 1.5-3, which are the SNR thresholds of most interest for analysis. We observe a consistent pattern that the stereographic and Mercator peak counts remain greater than the other peak counts, while the orthographic and gnomonic peak counts are closer to the spherical case, for opening angles less than 45$^{\circ}$ in the gnomonic case. The sine projection, in contrast, diverges from the spherical case at smaller opening angles, but is more accurate to the spherical case at greater opening angles. However, the gnomonic projection is unsuitable for opening angles between 90$^{\circ}$ and 270$^{\circ}$, as the gnomonic projection only applies to angles less than 45$^{\circ}$ which translates to a 90$^{\circ}$ opening angle as we define the masks.

\begin{figure*}
\centering
    \includegraphics[width=0.3\linewidth]{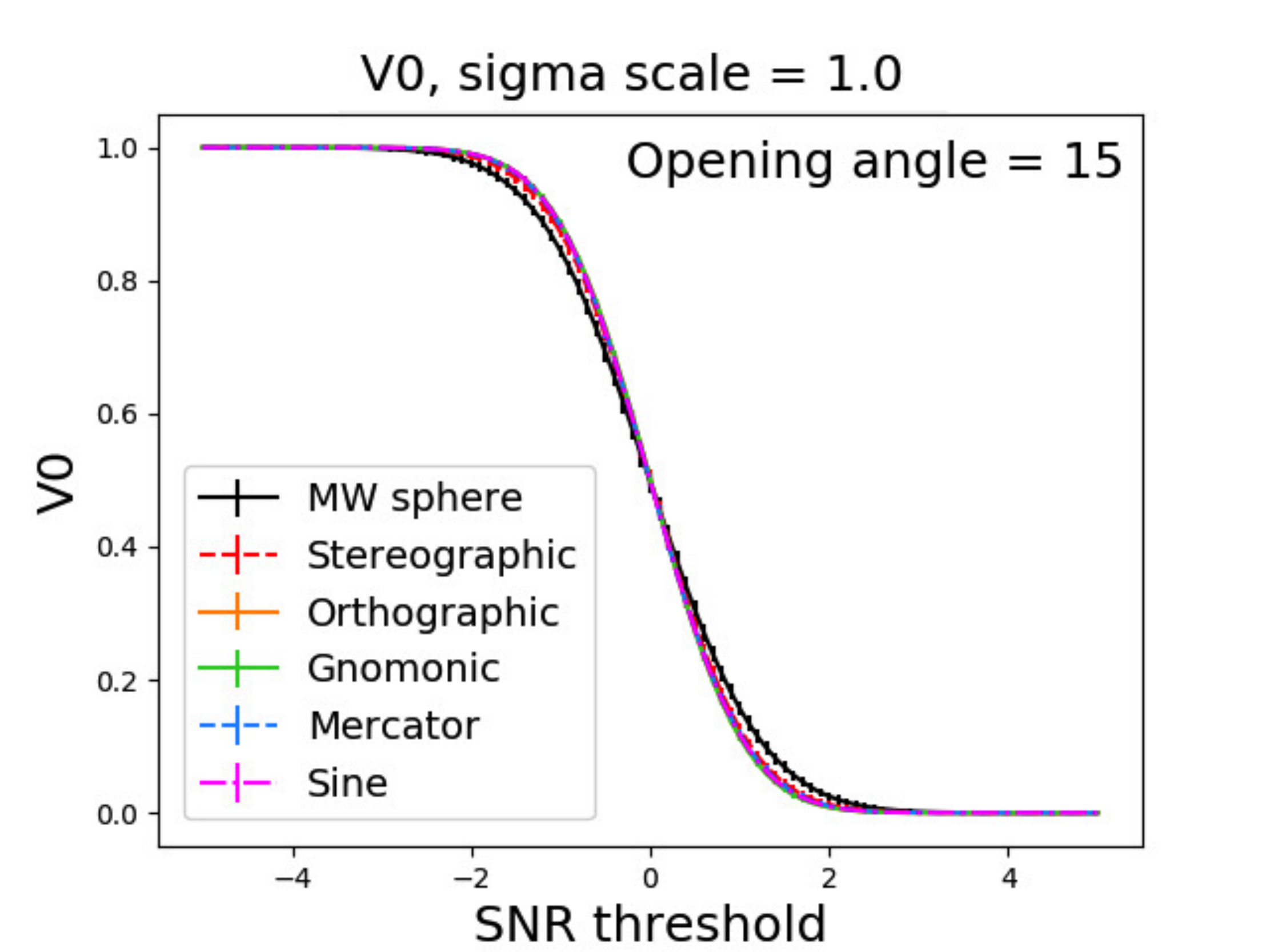}
    \includegraphics[width=0.3\linewidth]{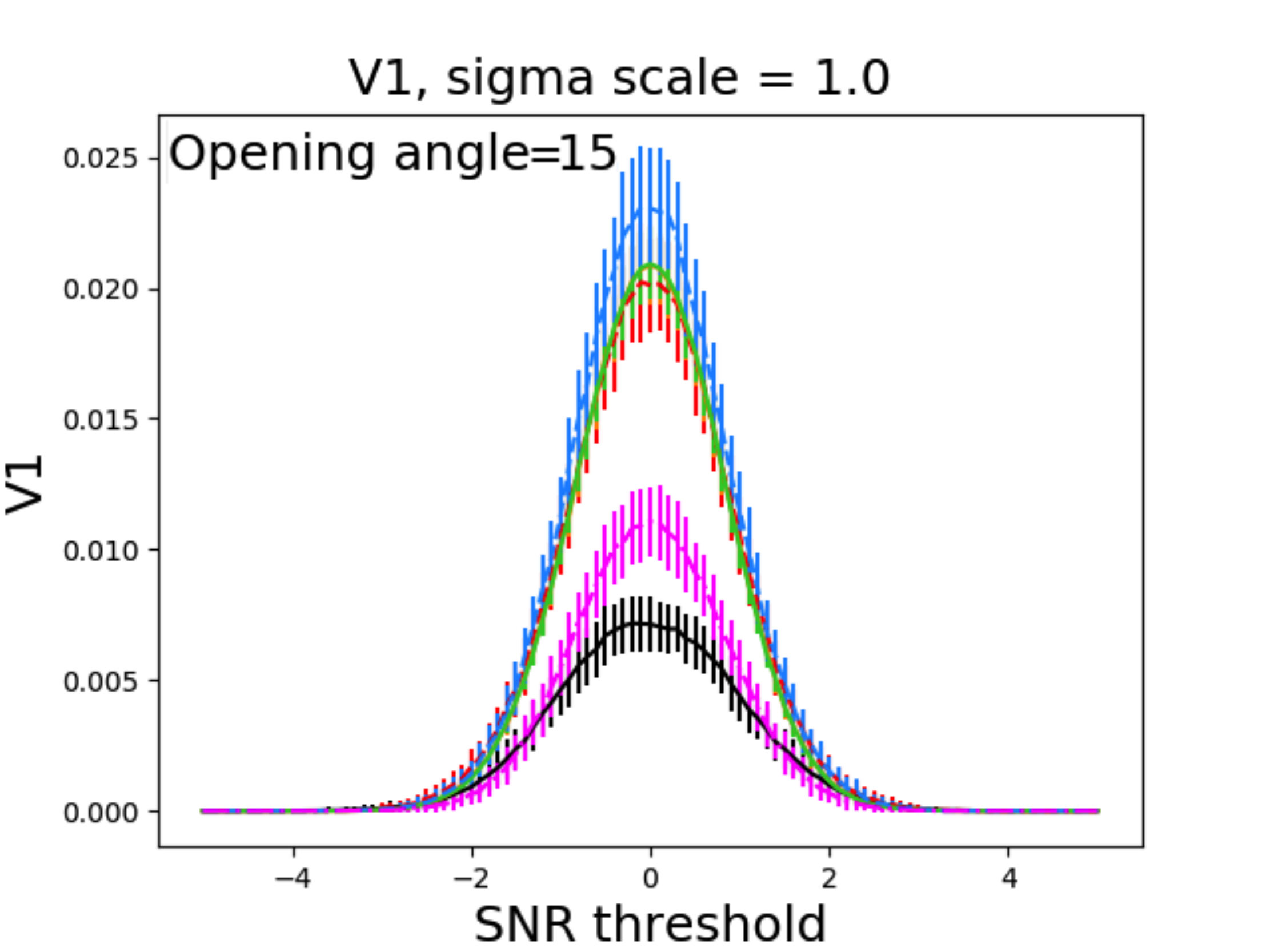}
	 \includegraphics[width=0.3\linewidth]{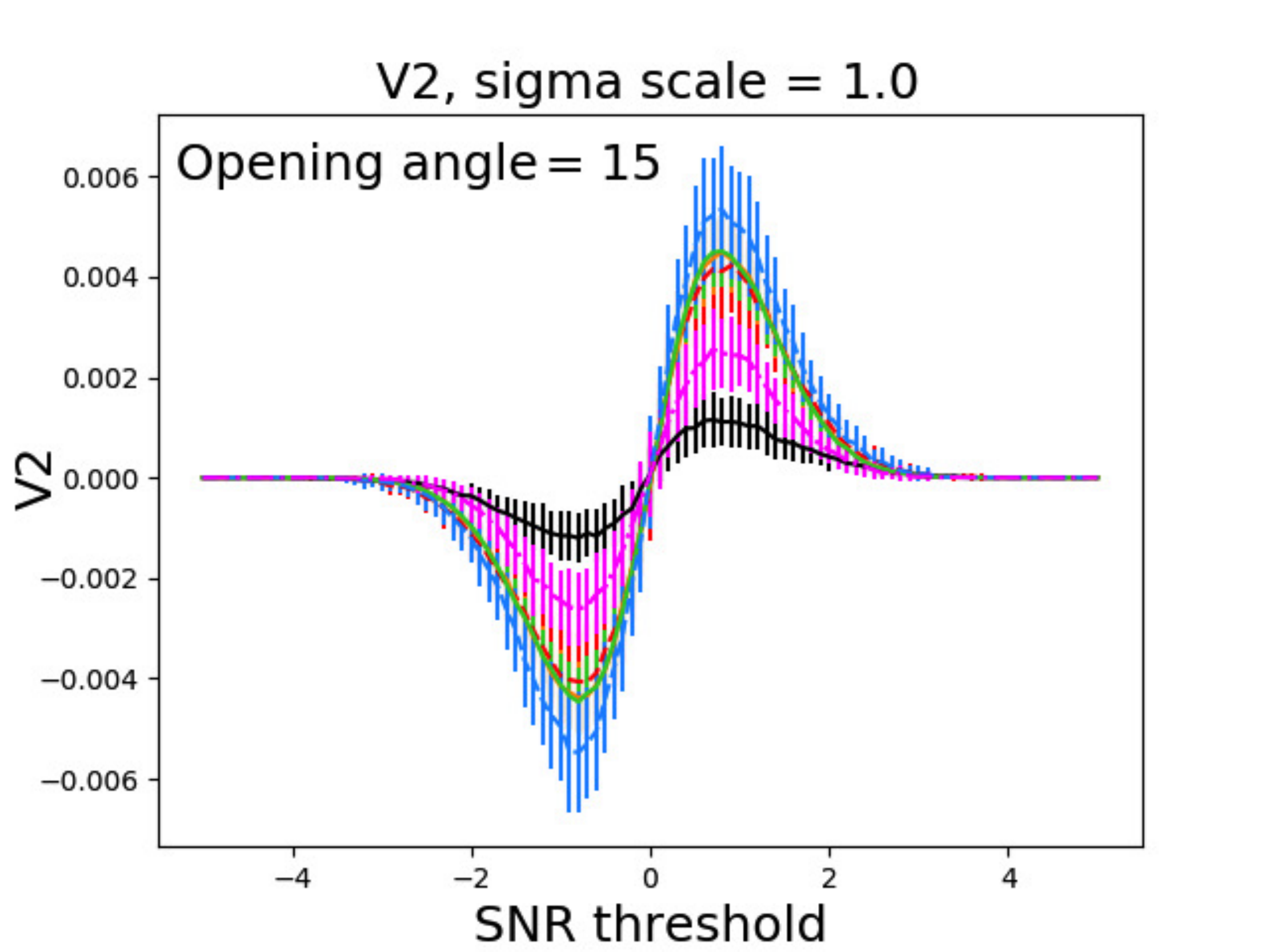} 
\par\medskip
    \includegraphics[width=0.3\linewidth]{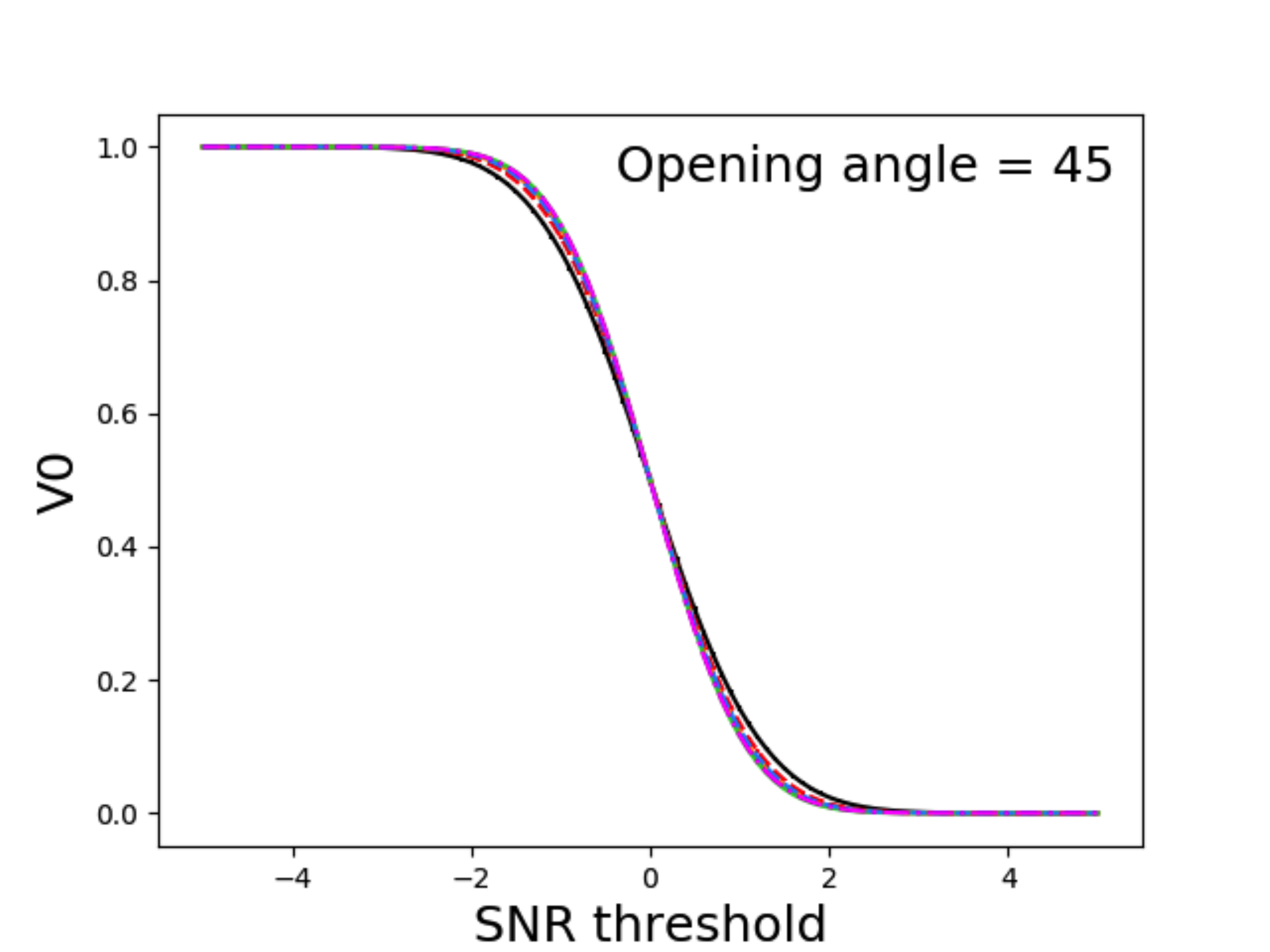} 
    \includegraphics[width=0.3\linewidth]{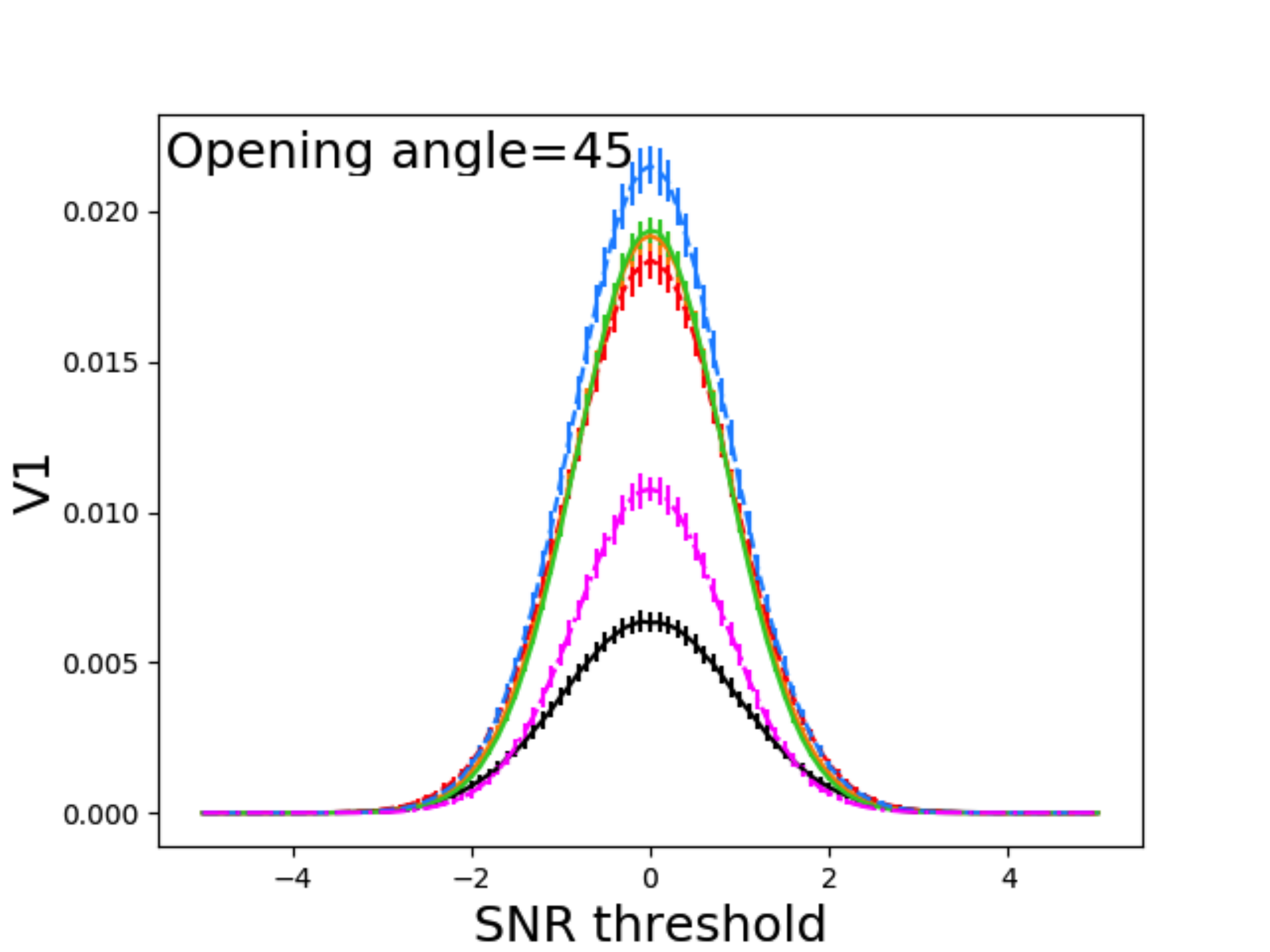}
    \includegraphics[width=0.3\linewidth]{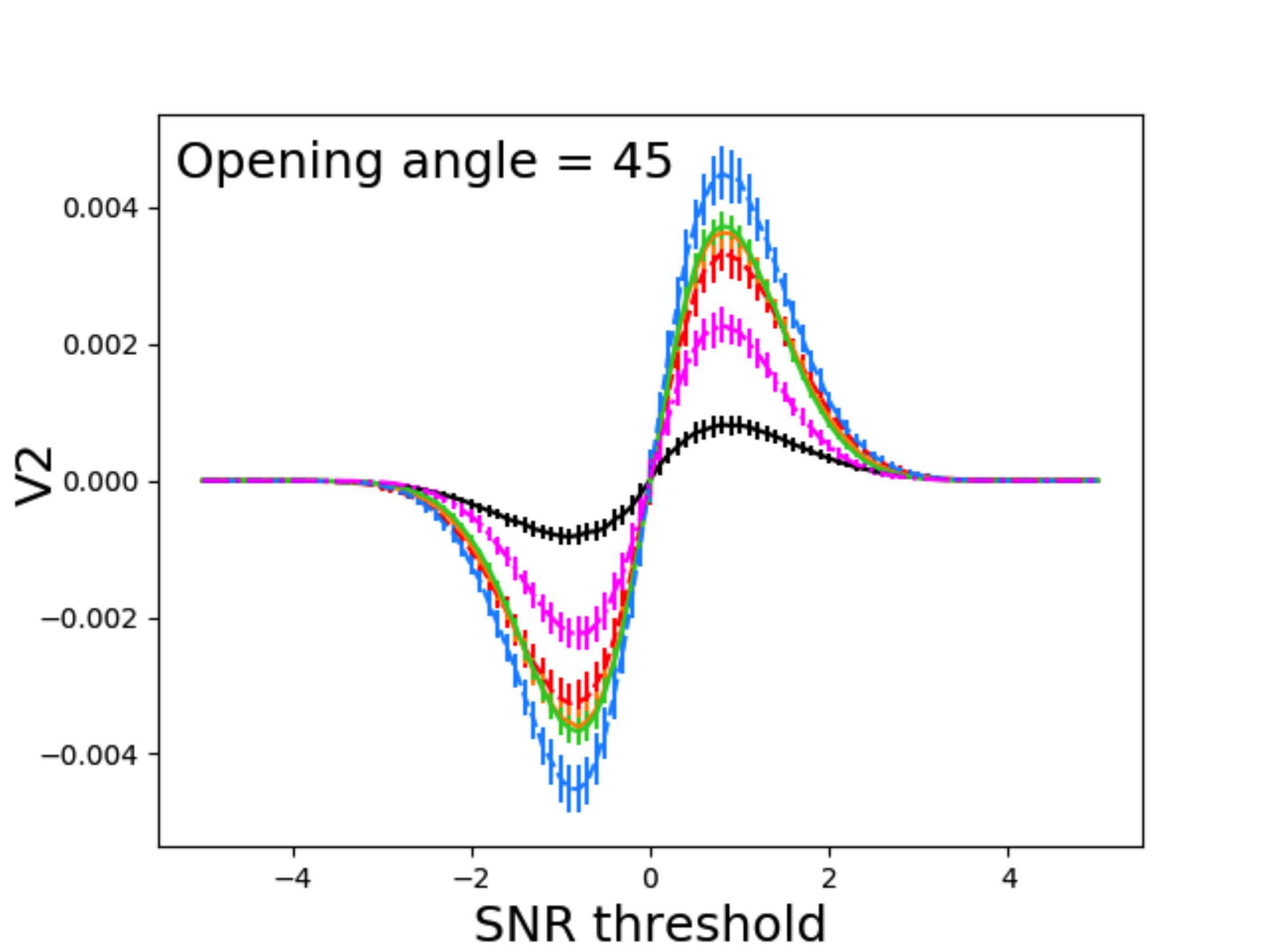}
\par\medskip
    \includegraphics[width=0.3\linewidth]{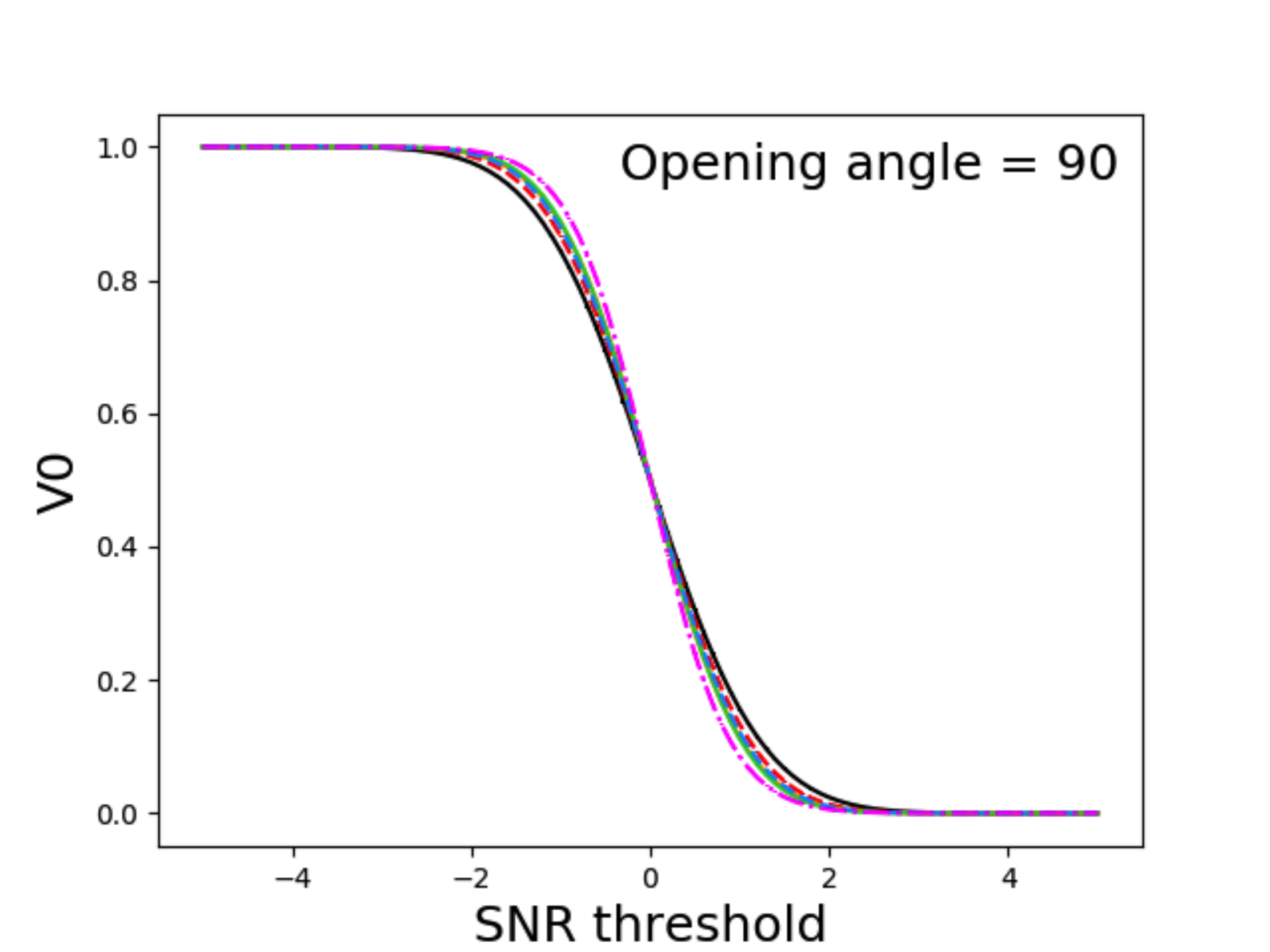}
    \includegraphics[width=0.3\linewidth]{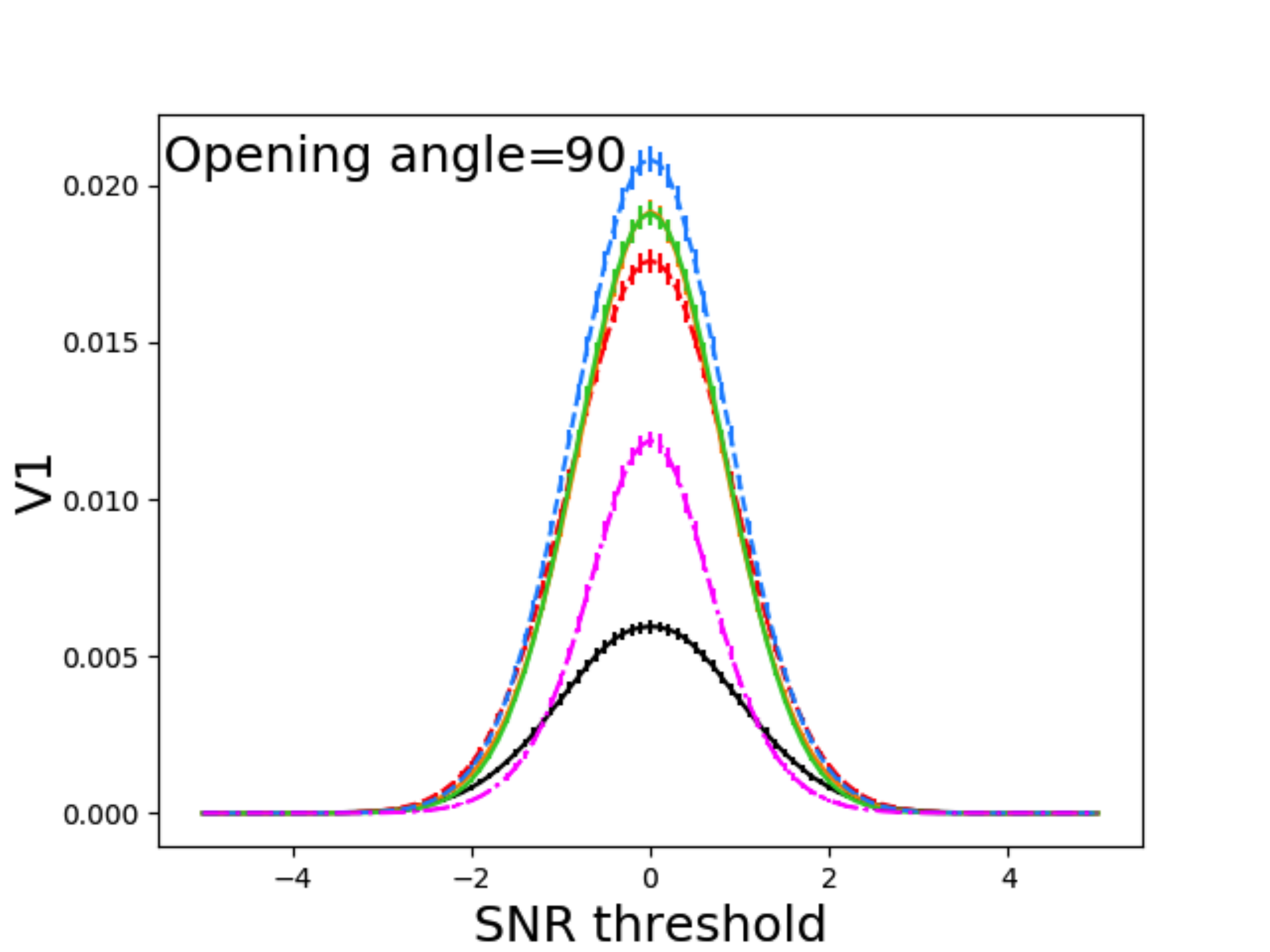}
    \includegraphics[width=0.3\linewidth]{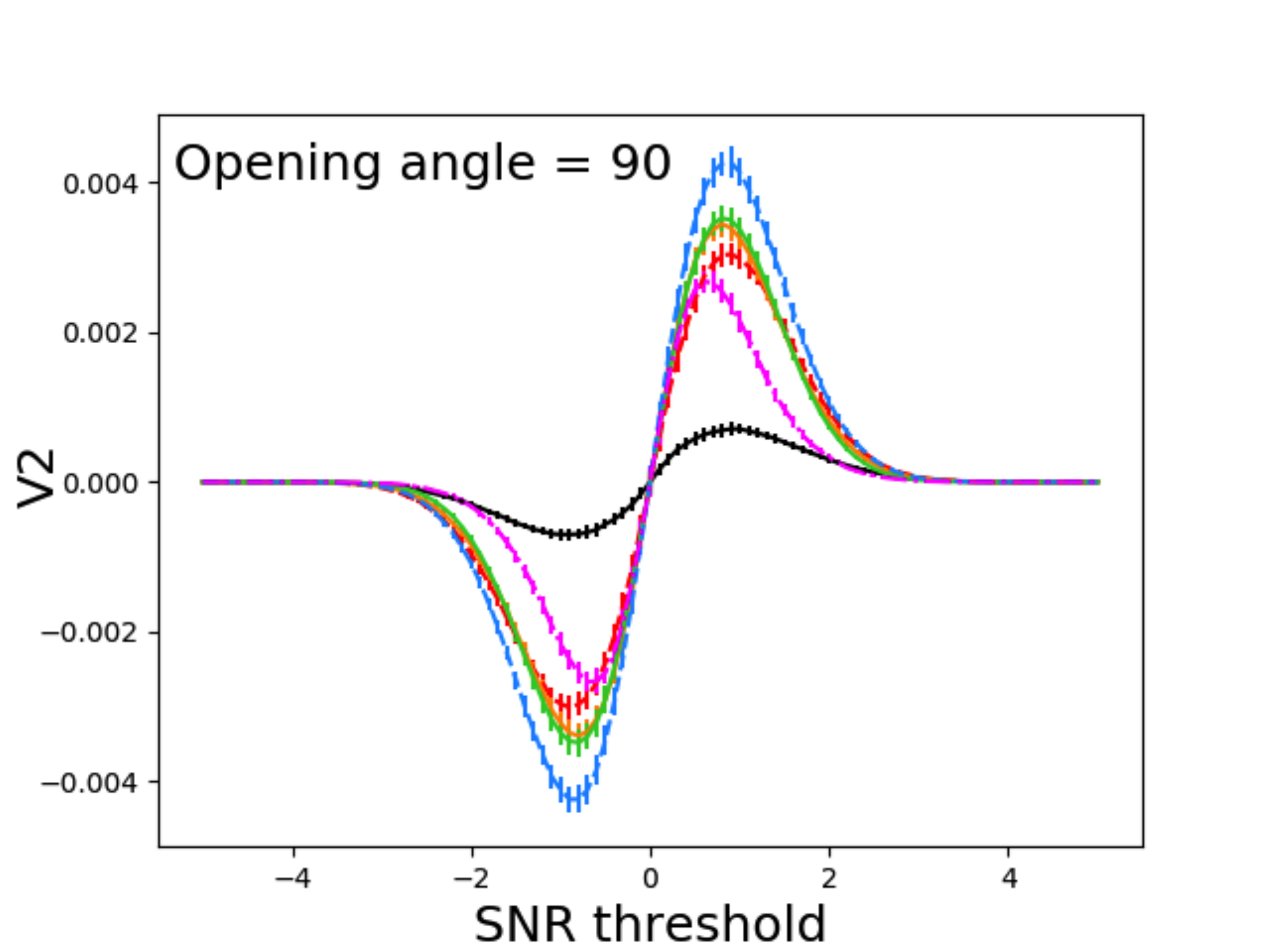}
\par\medskip
    \includegraphics[width=0.3\linewidth]{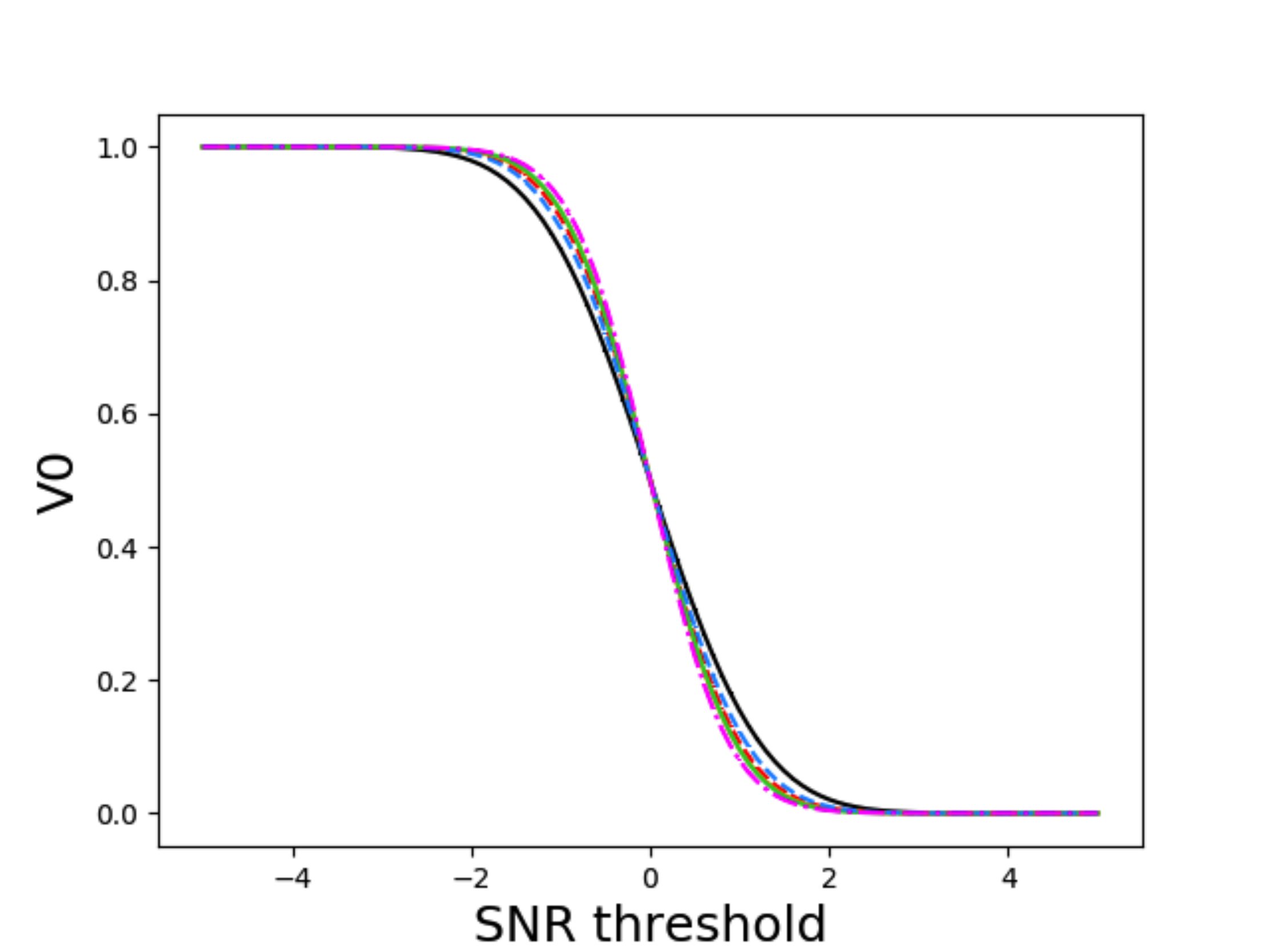}
    \includegraphics[width=0.3\linewidth]{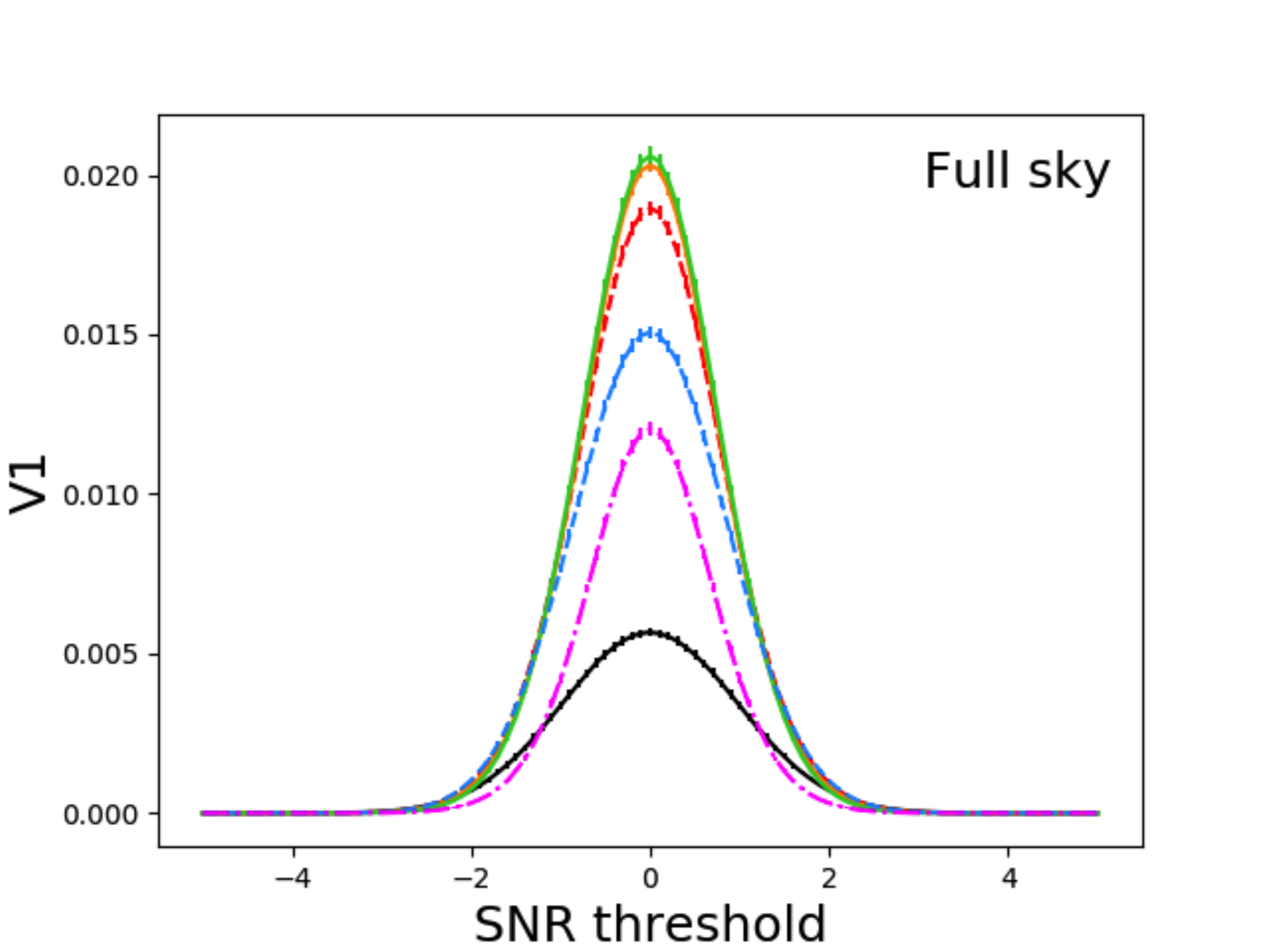}
    \includegraphics[width=0.3\linewidth]{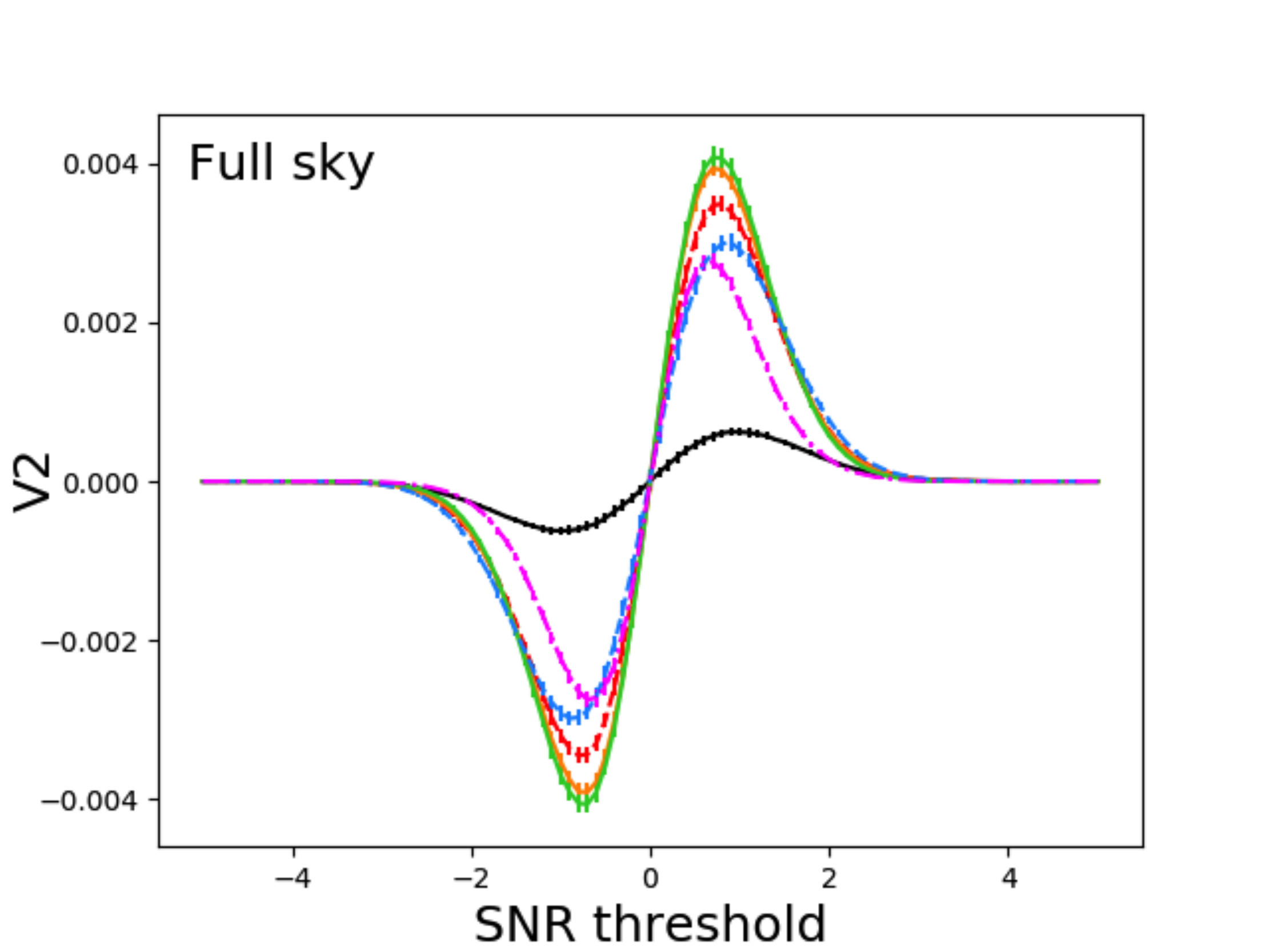}
\caption{Minkowski Functionals of the 2D field $V_0$, $V_1$ and $V_2$, with smoothing using $\sigma_s=1,5$ and opening angles 15$^{\circ}$, 45$^{\circ}$ and 90$^{\circ}$ for masked cases and the full sky case, at L=512 and for 100 iterations.}
\label{fig:MFs_sigma1}
\end{figure*}

\subsection{Minkowski functionals}
\label{sec:MFresults}

The Minkowski Functionals (MFs) described in equations \ref{eq:V0}, \ref{eq:V1}, \ref{eq:V2} for the spherical case and the five projections for L=512 with sigma scale $\sigma_s=1$ are displayed in Fig.~\ref{fig:MFs_sigma1} for the full sky case and 3 masked cases with opening angles 15$^{\circ}$, 45$^{\circ}$ and 90$^{\circ}$. We see that $V_0$ has the least difference between the spherical case and projected cases, while the MFs that incorporate the Dirac delta function and derivatives of the $_0\kappa$ maps -- $V_1$ and $V_2$ -- show a significant difference. This is primarily due to the Dirac delta function, which traces contours at a given threshold, that contributes most significantly to the differing values of the MFs. In maps with more noise, $V_1$ and $V_2$ are inflated due to false detections at the examined SNR thresholds. The resolution used also has a significant influence, as maps with a greater number of pixels resulted in lower values of $V_1$ and $V_2$. The projection method also affects the MFs through the presence of noise and distortion of shapes, which can be mitigated by using appropriate smoothing and high resolution. These factors heavily influence the contouring of the map, which impacts $V_1$ and $V_2$. However, we wish to identify and minimise other potential causes of the difference in MFs that are not caused by the projected geometry, and we discuss this below. We find that $V_0$ is not significantly affected by the projection method, so subsequent analysis will focus on $V_1$ and $V_2$. 

As with the peak counts, the degree of smoothing significantly impacts the MFs $V_1$ and $V_2$. When the smoothing is not sufficient, significant noise may still be present in the SNR map which leads to amplified values for $V_1$ and $V_2$. We find that the presence of noise has a greater impact on the MFs $V_1$ and $V_2$ than the choice of projection. The projected maps have significantly greater noise than the map on the sphere due to how the smoothing is performed. We stress the importance of accounting for noise and using appropriate smoothing. Further discussion of the effect of smoothing is detailed in \ref{appendix:smoothing}.

We expect to find that on small scales projection effects are minimised and the projected MFs will more closely match the spherical MFs. However, we find that the effects of pixelisation and smoothing have a significant impact on the MFs $V_1$ and $V_2$ for different projections for all opening angles of observed area. Using smaller opening angles leads to significantly greater error, as seen in Fig.~\ref{fig:MFs_sigma1}, which is caused by the reduced number of pixels observed, hence pixelisation effects are more prominent. The effect of the number of pixels used to calculate MFs is significant enough to dominate over any distortion caused by projection. However, even with large errors it is still clear the manner in which projection affects the MFs on different scales. This can be mitigated by increasing the number of iterations or using greater resolution to achieve higher pixel count. 

One possible solution to this issue is to use projection-dependent smoothing i.e. applying a different smoothing scale for each projection in order to produce SNR maps that are more closely matched to the spherical case for small opening angles. The MFs measure the areas and contours of the map and not individual peaks, so increased smoothing does not remove significant information as it does with the peak counts. Under the assumption that there will be no projection effects on very small scales, we evaluate the maximum values of $V_1$ across a range of $\sigma_s$, as shown in Fig.~\ref{fig:maxV1_varysigma} in Appendix \ref{appendix:smoothing}, and for each projection we select a $\sigma_s$ that corresponds to a value of $V_1$ closest to a selected $V_1$ maximum for the spherical case on small scales. The MF $V_2$ is slightly less affected by $\sigma_s$ used than $V_1$, so we use $V_1$ to define the projection-dependent smoothing. 

\begin{figure*}
\centering
    \includegraphics[width=0.3\linewidth]{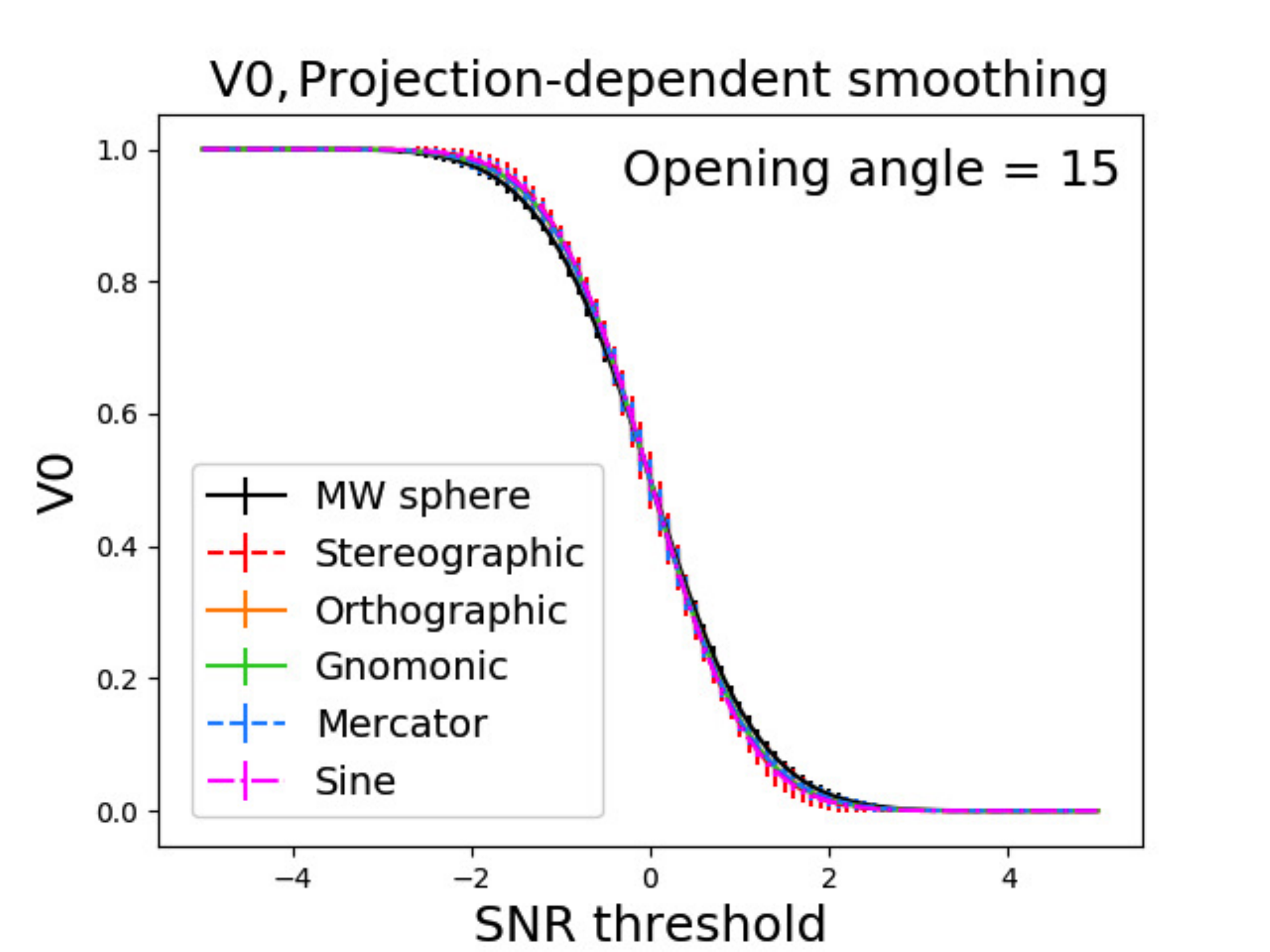}
    \includegraphics[width=0.3\linewidth]{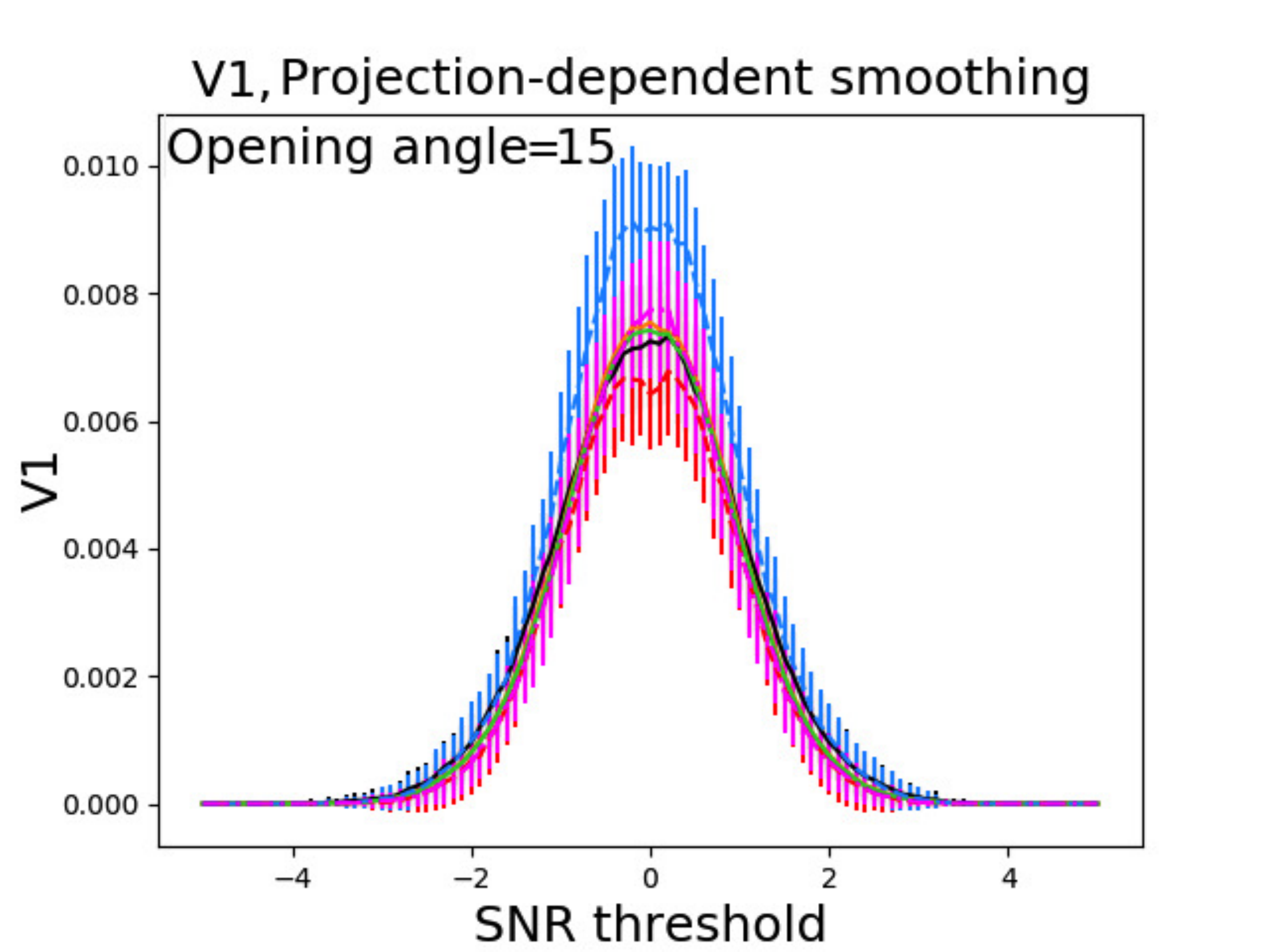}
	 \includegraphics[width=0.3\linewidth]{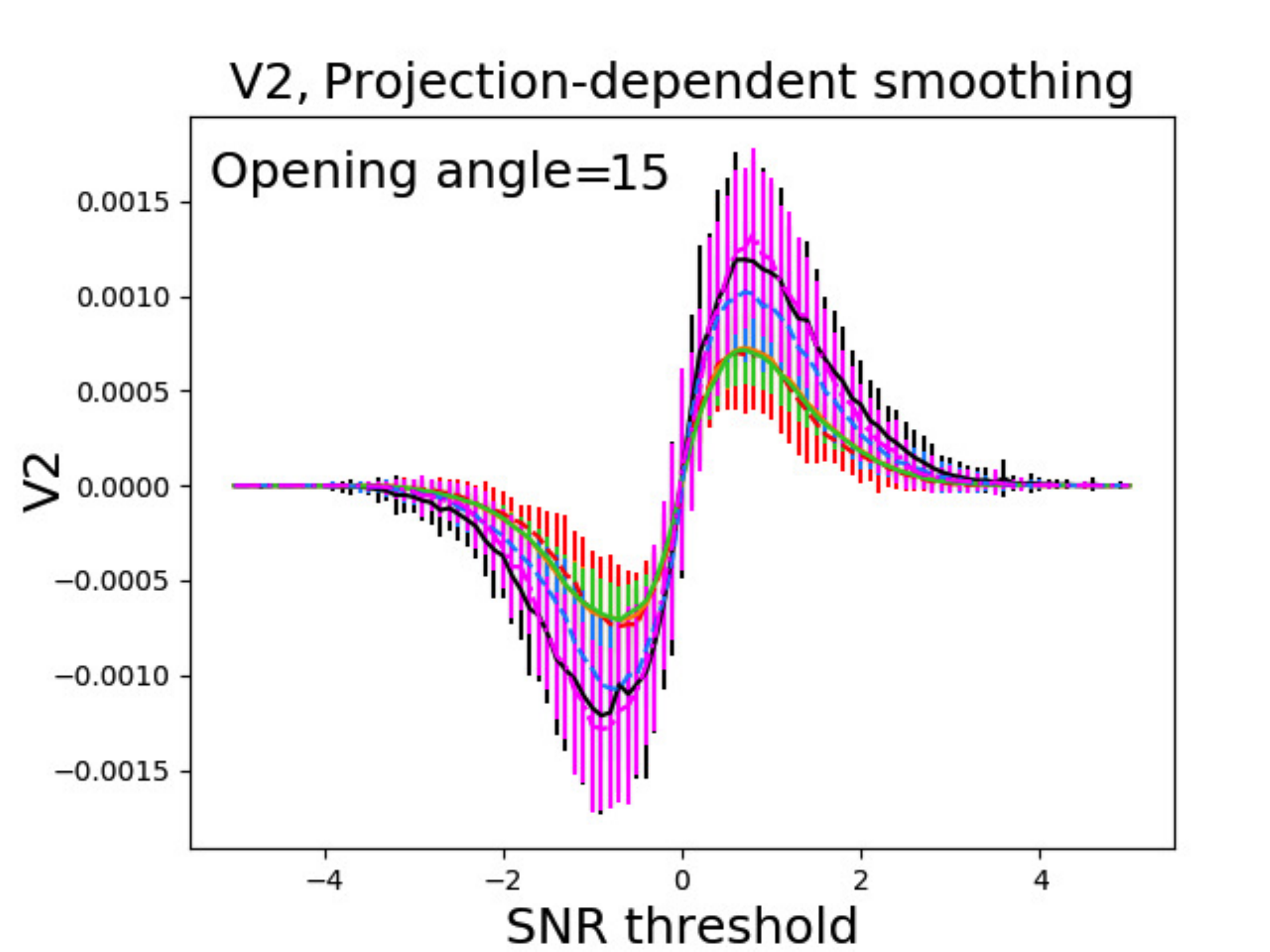} 
\par\medskip
    \includegraphics[width=0.3\linewidth]{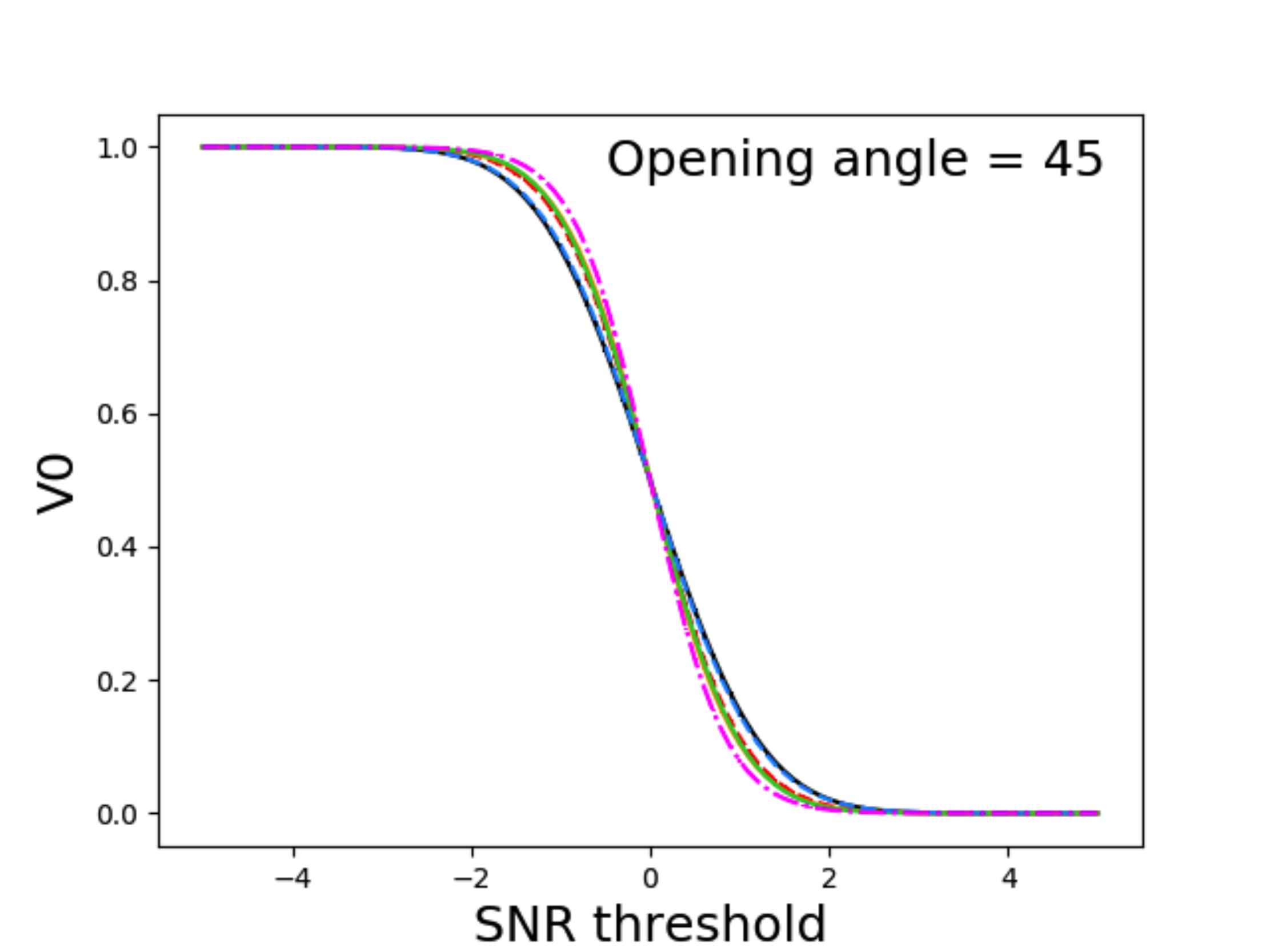} 
    \includegraphics[width=0.3\linewidth]{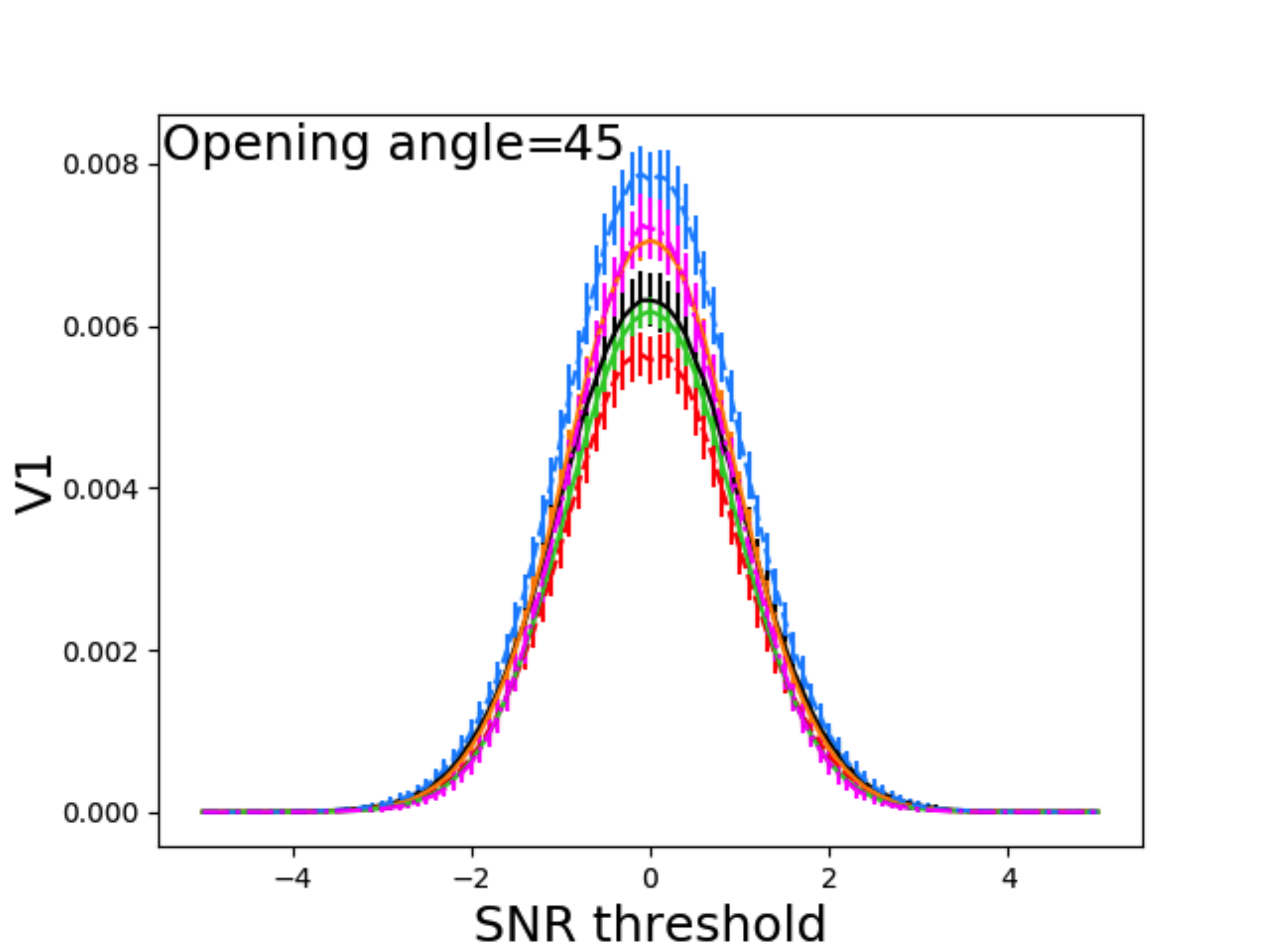}
    \includegraphics[width=0.3\linewidth]{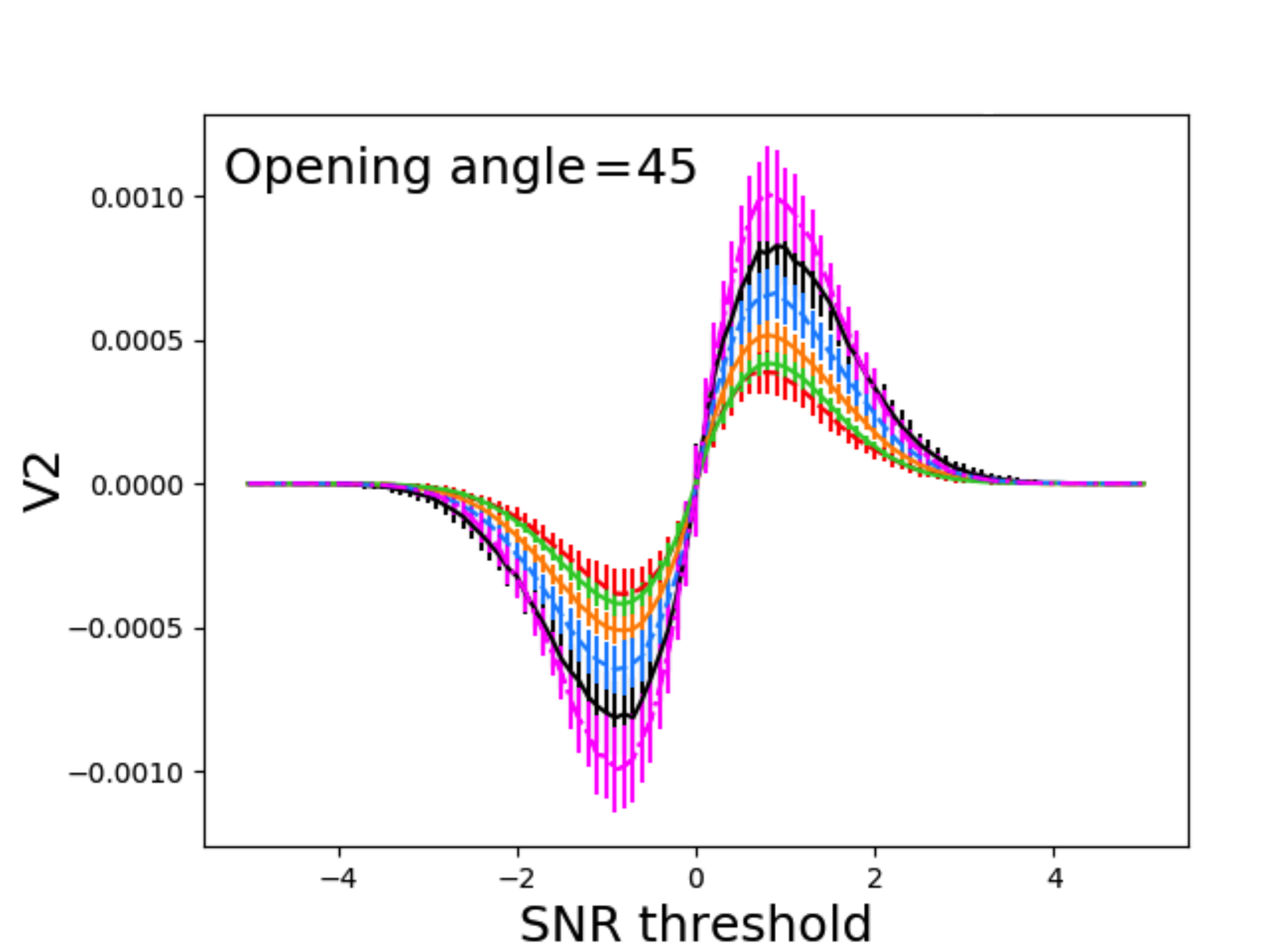}
\par\medskip
    \includegraphics[width=0.3\linewidth]{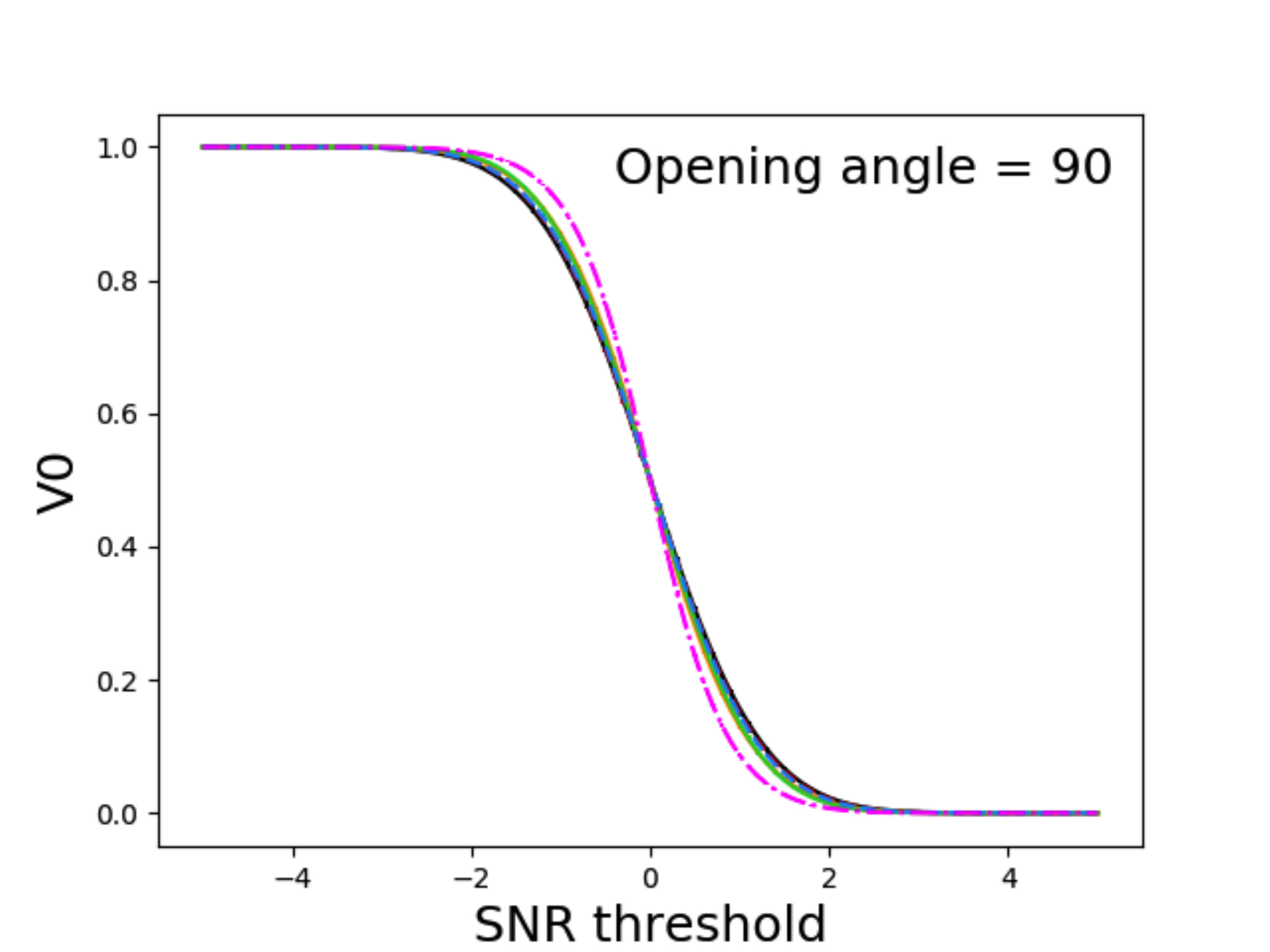}
    \includegraphics[width=0.3\linewidth]{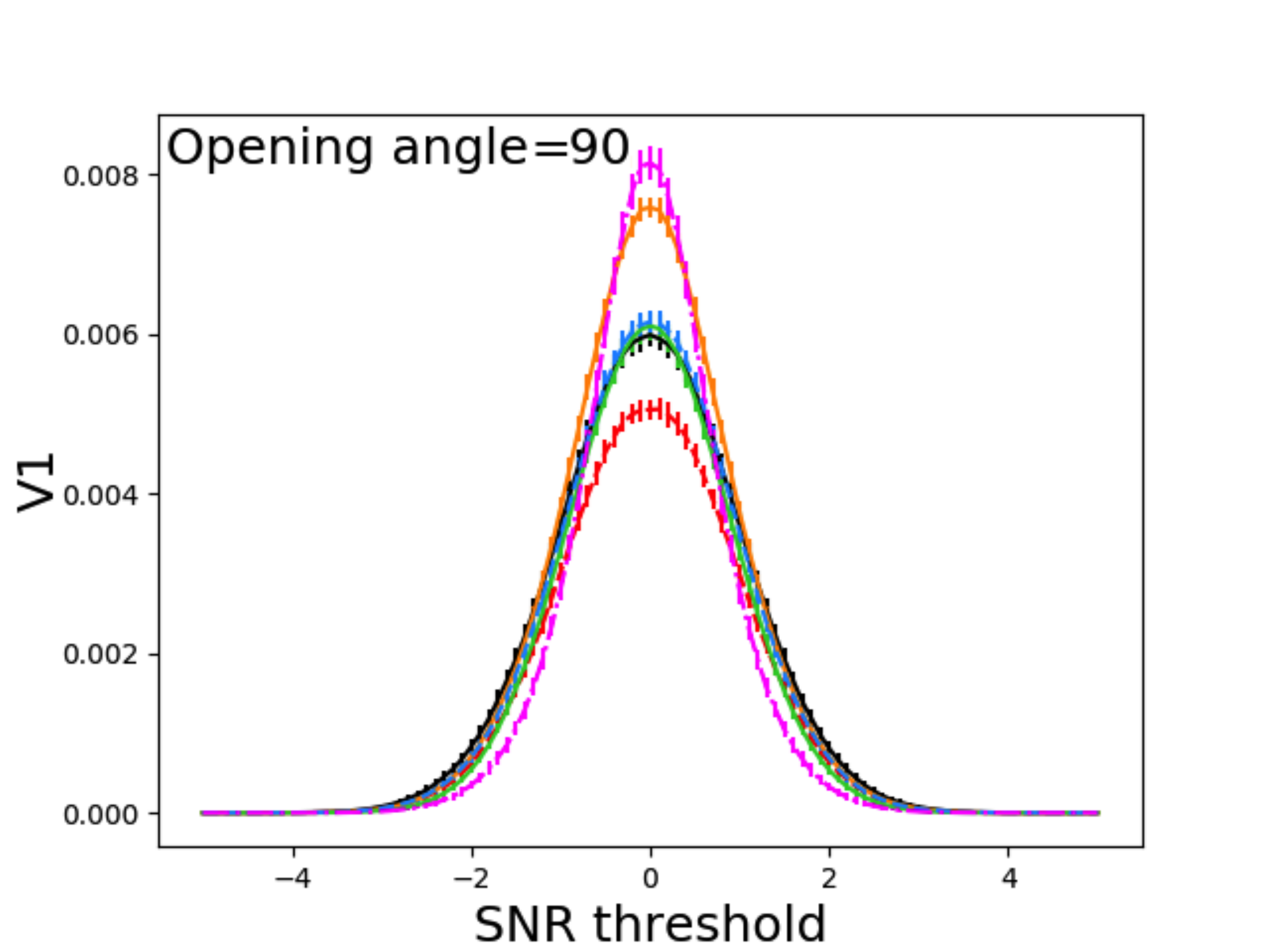}
    \includegraphics[width=0.3\linewidth]{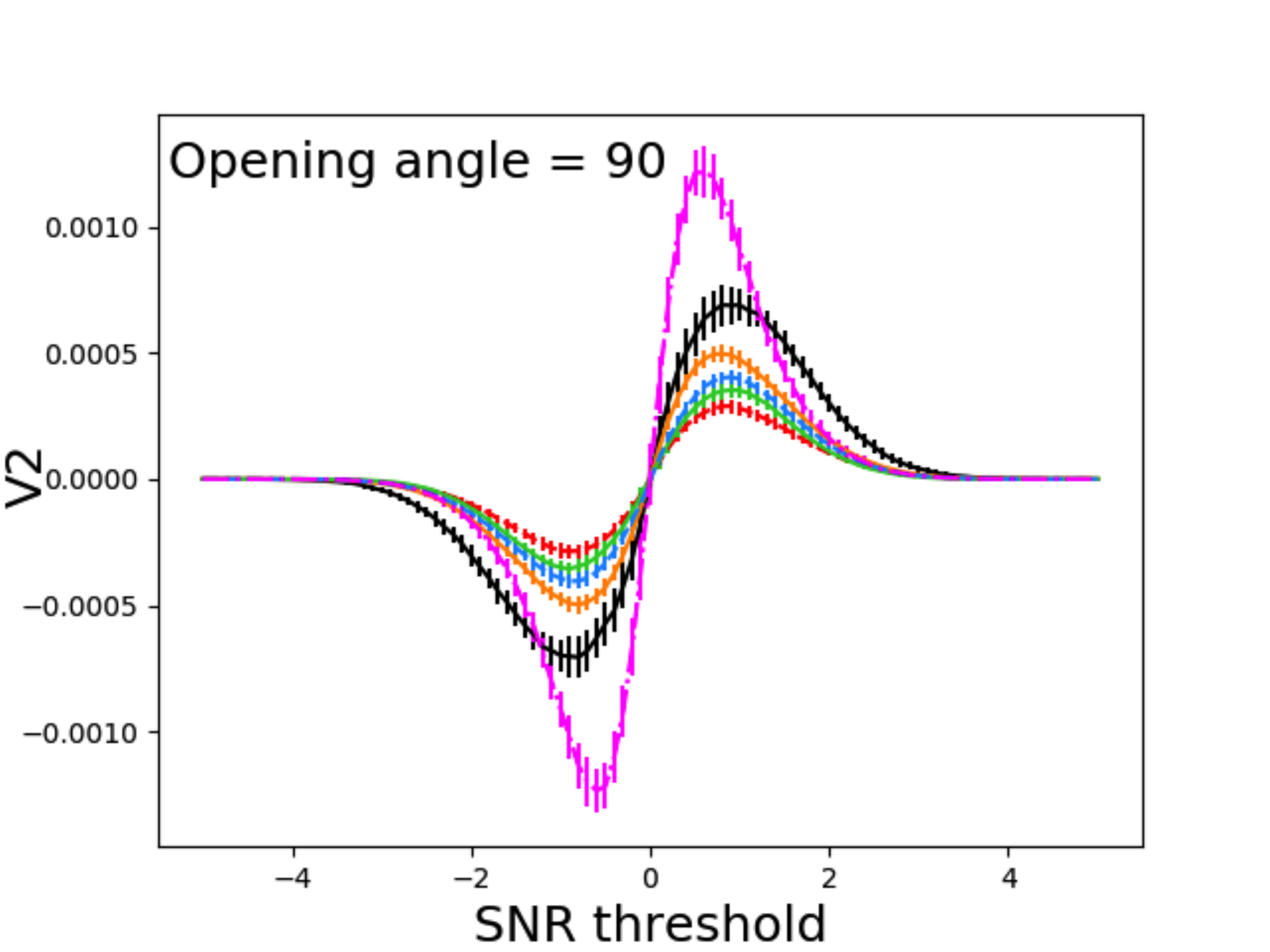}
\par\medskip
    \includegraphics[width=0.3\linewidth]{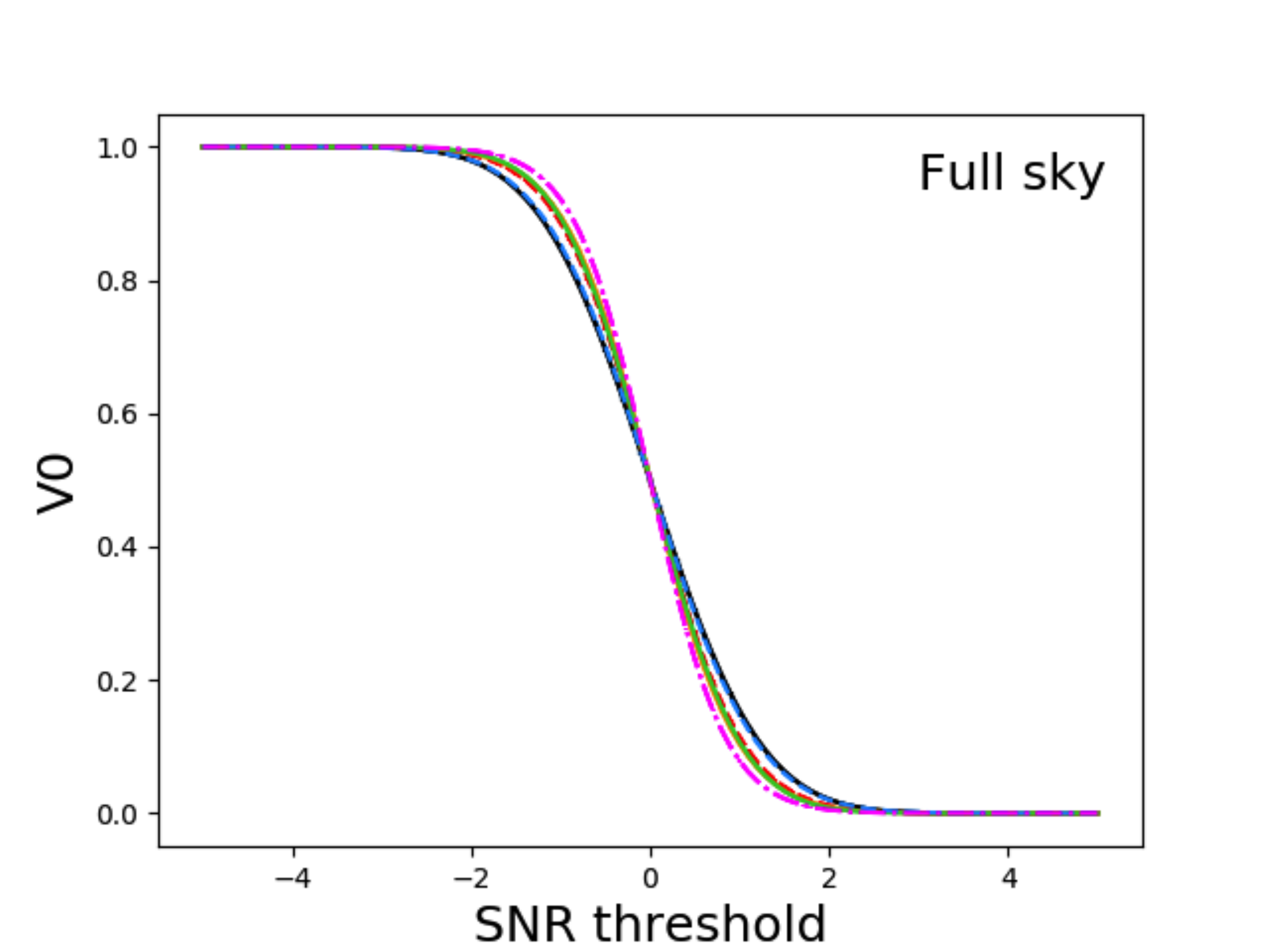}
    \includegraphics[width=0.3\linewidth]{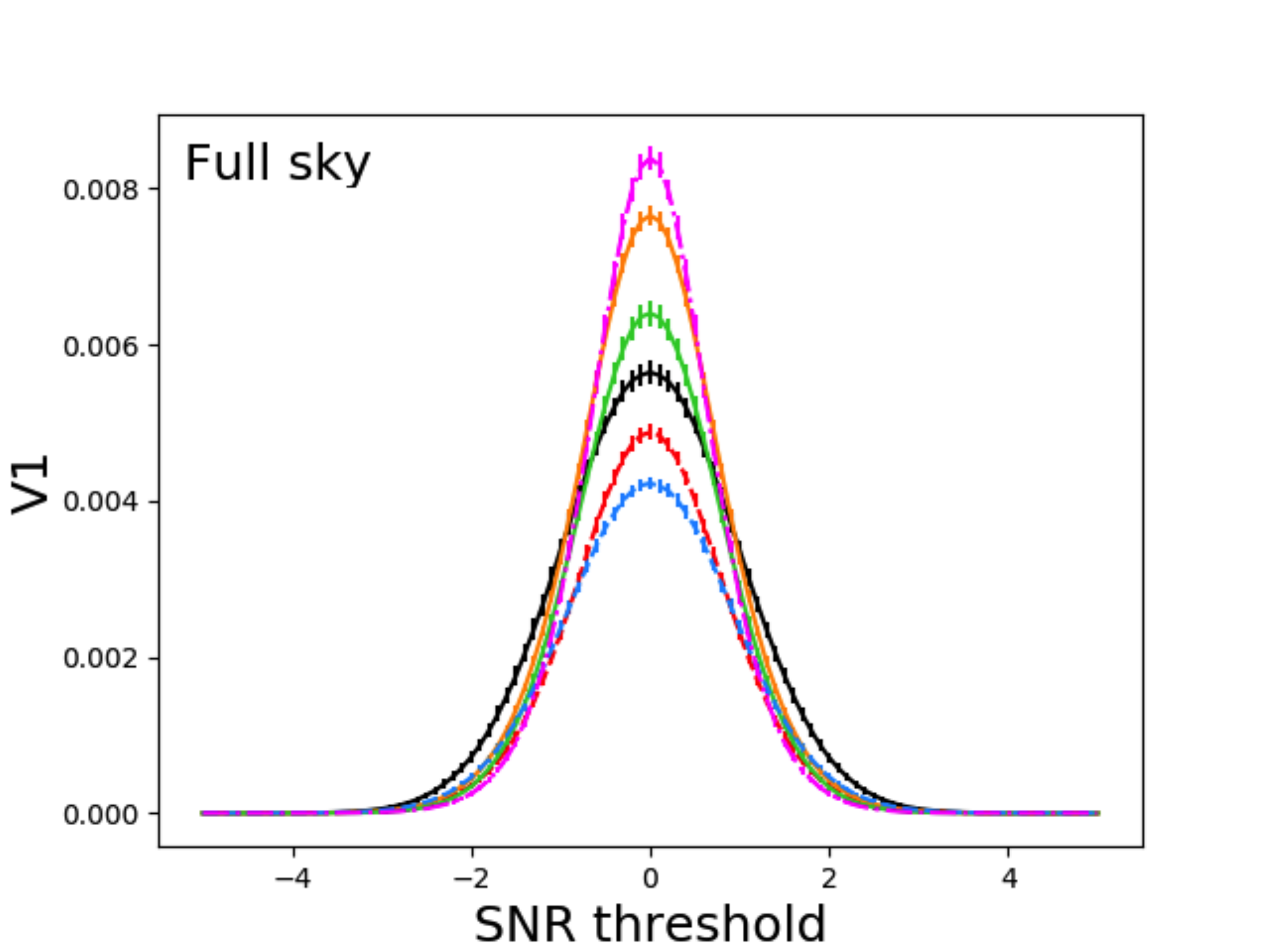}
    \includegraphics[width=0.3\linewidth]{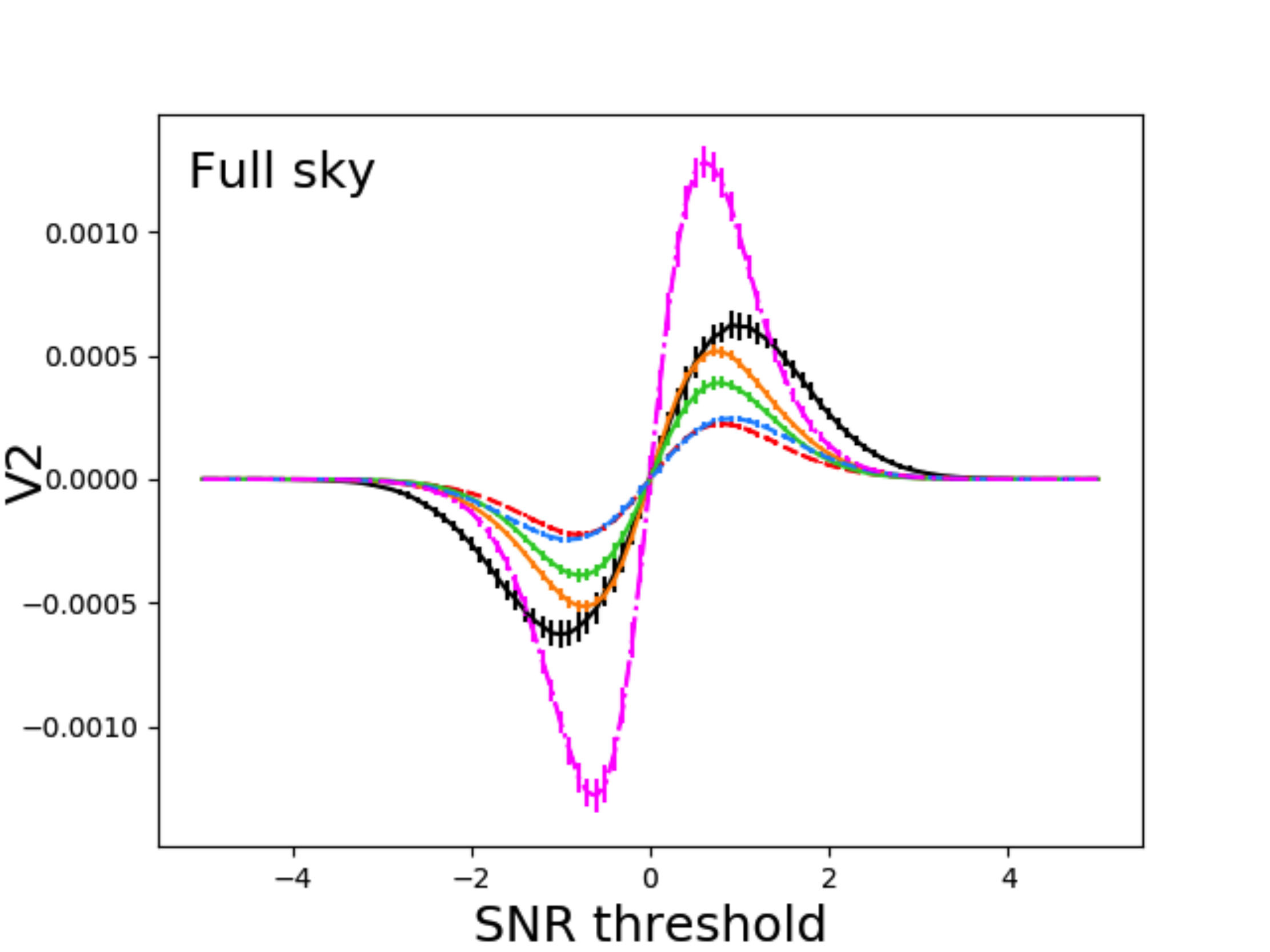}
\caption{Minkowski Functionals of the 2D field $V_0$, $V_1$ and $V_2$, using projection-dependent smoothing for opening angles 15$^{\circ}$, 45$^{\circ}$ and 90$^{\circ}$ for masked cases and the full sky case, at L=512 and for 100 iterations. We define smoothing scale for each case to be $\sigma_s=1$ for the spherical case, $\sigma_s=10$ for the stereographic projection, $\sigma_s=4$ for the orthographic projection, $\sigma_s=4$ for the gnomonic projection, $\sigma_s=8$ for the mercator projection and $\sigma_s=2$ for the sine projection}
\label{fig:MFs_variablesigma}
\end{figure*}

The MFs obtained using projection-dependent smoothing are displayed in Fig.~\ref{fig:MFs_variablesigma} for the masked case with opening angle 15$^{\circ}$ and the full-sky case. With the projection-dependent smoothing, the MFs of the full-sky case should more accurately reflect the influence of the projections not caused by noise. We see examples of this displayed in Fig.~\ref{fig:MFs_variablesigma} for the masked case with opening angle 15$^{\circ}$ and for the full sky case. We observe that using projection-dependent smoothing causes the small-angle projected MFs to more closely align with the spherical case. However, in many cases the projected MFs, while matching the spherical case for small angles, are now reduced below the spherical MFs on larger angles. In the full-sky case, although it is not possible to separate the effect of increased smoothing from the influence of shape distortions from projection, using the assumption that the smoothing used produces equivalence for the MFs on small scales, the resulting MFs should approximately reflect the effect of each projection separated from the effects of noise. Since the MFs do not match the spherical case exactly, even with projection-dependent smoothing, analysis of each MF should be performed separately for each projection.

\section{Conclusions}
\label{sec:conclusion}

Whenever a planar approximation is made to the spherical celestial sphere a projection must be chosen. Here we investigate the effect of five different projections on the peak count statistic and Minkowski functionals of weak lensing convergence maps. 

We use the software packages \texttt{SSHT} and \texttt{massmappy} to generate simulated recovered convergence maps. This is done using the following perscription; first we simulate shear maps on the sphere, project these shear maps to the plane, reconstruct the convergence maps natively on the plane, construct the SNR map from the convergence map, and identify peaks for a range of Signal-to-Noise Ratio (SNR) thresholds to obtain peak count statistics. 

In the five examined projections (see Section \ref{sec:proj} for details on the projections) the peak counts derived from planar projected maps are greater than peak counts for the spherical case at low SNR thresholds and lower than the spherical case for high SNR thresholds; provided that the peak count numbers are adjusted for the differing number of pixels in the map and area of sky covered. We find that all of the examined projections have drawbacks in comparison to peak counts evaluated on the sphere. The projected peak counts consistently have lower maximum SNR thresholds than the spherical case. The orthographic projection produces peak counts that are most similar to the spherical case for high SNR thresholds when accounting for differences in pixelisation. While the peak counts from the orthographic and gnomonic projections  most closely match the spherical case over the greatest range of SNR thresholds, the peak counts are still overestimated at low SNR thresholds and underestimated at high SNR thresholds. While increasing the smoothing and reducing observed area improves the peak count accuracy, the projected peak counts cannot be made to match the spherical peak counts at all SNR thresholds. Further smoothing to further reduce noise would result in the obscuration of real peaks, and a loss of signal. Thus we recommend that peak count analysis of large areas of sky be performed on the sphere if possible.

We also evaluated the Minkowski Functionals on the five projections in comparison to the spherical case. Assuming the same parameters used in the projections as the spherical case, we found large differences in $V_1$ and $V_2$ between the spherical case and the projected cases. This difference is primarily due to the greater noise in the maps reconstructed after projection caused by smoothing being applied unevenly (locally asymmetrical compared to the pixel coordinate system) on the projected map in comparison to the evenly applied smoothing on the sphere. This noise effect is significantly greater than other projection effects such as the distortion of shapes from the projection and must be accounted for when evaluating MFs on projections. One possible method of reducing the influence of noise is to apply projection-dependent smoothing, by establishing a smoothing scale for which the MFs match the spherical case on small areas of sky (where the influence of projection is minimised), and subsequently applying these smoothing scales to the projected data for larger areas. 

While we find that while the projected peak counts, and MFs, resemble the spherical values at small sky areas of sky coverage, as the sky coverage increases they diverge. Planar projections remain appropriate for analyses over small areas of sky, but for future experiments with greater sky coverage the approximations break down and we caution against using the planar projection for analysis and emphasise the necessity of moving towards analysis on the sphere. We must conclude that care must be taken when using projected MFs and peak statistics, and comparisons to simulated fields should be done in the same geometry as the observations.

\section*{Acknowledgements}
We thank Jason McEwen and Paniez Paykari for valued input and discussion. We acknowledge the funding and support provided by the Science \& Technology Facilities Council (STFC). TK is supported by a Royal Society University Research Fellowship.




\bibliography{bibliography.bib} 



\newpage

\appendix

\section{Smoothing scale for reconstruction}
\label{appendix:smoothing} 

We used a smoothing parameter $s=20\text{arcmin}=2\times20.0\times\pi/(60\times 180\times 2.355)\times\sigma_s$ for both the peak counts and Minkowski functional analysis on the sphere. Due to the differences in resolution between the spherical case and projected cases the smoothing parameter needs to be adjusted for each projection to ensure that the degree of smoothing applied produces an equivalent mass map. 

The MFs are significantly influenced by noise in a different manner to the peak counts because $V_1$ and $V_2$ measure the contours of a 2D map which are distorted by projection, hence the degree of smoothing must be carefully considered for the MFs and handled separately to smoothing for peak counts. Increasing smoothing tends to decrease the maximum value of $V_1$ and $V_2$ due to reduced noise, bringing several projection MFs closer to the MFs measured in the spherical case. Using smaller opening angles for observed area of sky increases in error on the MFs, but there is no significant difference in the mean of the shape of observed MFs $V_1$ and $V_2$ for low smoothing with $\sigma_s=1$, since the projected map MFs are dominated by noise regardless of observed area.

\begin{figure*}
\centering
    \includegraphics[width=0.8\linewidth]{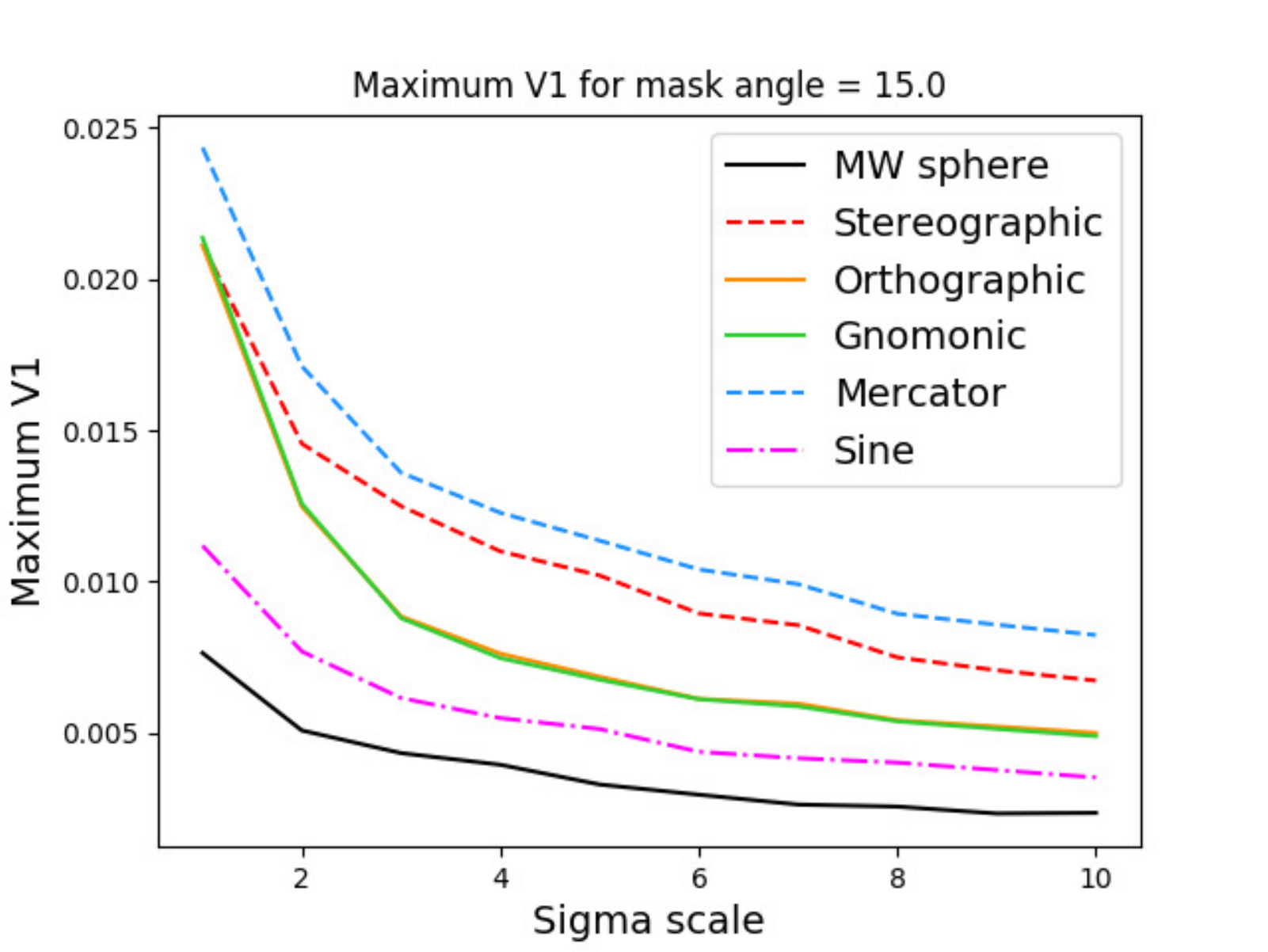}
\caption{Maximum value of Minkowski Functional $V_1$ over a range of smoothing scales $\sigma_s$ for window with opening angle 15$^{\circ}$ and 20 iterations.}
\label{fig:maxV1_varysigma}
\end{figure*}

In Fig.~\ref{fig:maxV1_varysigma} we display the maximum value of the MF $V_1$ as a function of smoothing scaling parameter $\sigma_s$ for the masked case with a window of opening angle 15$^{\circ}$. The maximum $V_1$ values decrease at a similar rate across the projections and spherical case, with little overlap other than the orthographic and gnomonic cases. While the maximum $V_1$ values consistently decrease, they eventually reach a plateau and do not converge even for high $\sigma_s$. A similar pattern is observed with the maximum and minimum values of $V_2$. This implies that even for high smoothing the spherical and projected MFs will not converge, hence there is an inherent difference in the MFs in each geometry after the noise is removed even for small opening angles. In the small area case, where we expect projection effects to be minimised, we still do not find convergence and is suspected to be due to pixelisation effects due to lower relative resolution, assuming both the small angle and full sky case use the same maximum bandlimit. 

However we can define an equivalent smoothing for each projection that would produce the same MF as the spherical case in the small opening angle limit i.e. a projection dependent smoothing. We use Fig.~\ref{fig:maxV1_varysigma} to identify an appropriate $\sigma_s$ for each projection to match $\sigma_s=1$ for the spherical case in order to perform projection-dependent smoothing. In this paper, we select the sphere smoothing as the standard $\sigma_s=1$, and then use $\sigma_s=10$ for the stereographic projection, $\sigma_s=4$ for the orthographic projection, $\sigma_s=4$ for the gnomonic projection, $\sigma_s=8$ for the Mercator projection and $\sigma_s=2$ for the sine projection.



\label{lastpage}
\end{document}